\documentclass[aps,reprint,prb,twocolumn]{revtex4-1}
\pdfoutput=1

\usepackage{graphicx}
\usepackage{amssymb}
\usepackage{amsmath}
\usepackage{amsthm}
\usepackage{mathtools}
\usepackage{textcomp,gensymb}
\usepackage{slashed}
\usepackage[colorlinks=true, linkcolor=blue, citecolor=blue, urlcolor=blue]{hyperref}
\usepackage{color,rotating}
\usepackage{bm}
\usepackage[normalem]{ulem}
\usepackage{bbm}
\usepackage{array}
\usepackage{pstricks}
\usepackage{pstricks-add}
\usepackage{colonequals}
\usepackage{tabularx}
\usepackage[utf8]{inputenc}

 \graphicspath{{./figures/}}

\newcommand{\IR}{{\small IR}}
\newcommand{\FRG}{{\small FRG}}
\newcommand{\RG}{{\small RG}}

\newcommand{\STr}{\operatorname{STr}}
\newcommand{\LPA}{{\small LPA}}

\newcommand{\CPP}{{\small C++}}
\newcommand{\NJL}{{\small NJL}}
\newcommand{\eg}{{\textit{e.g.}}}
\newcommand{\ie}{{\textit{i.e.}}}
\newcommand{\cf}{{\textit{cf.}}}

\newcommand{\fpGN}{{\small GN}}
\newcommand{\fpTh}{{\small Th}}
\newcommand{\fpB}{{\small B}}

\newcommand{\Eqref}[1]{Eq.~\eqref{#1}}
\newcommand{\mc}[1]{\mathcal{#1}}
\newcommand{\lmat}{\left( \begin{matrix}}
\newcommand{\rmat}{\end{matrix} \right)}
\renewcommand{\d}{\mathrm{d}}
\newcommand{\T}{\mathsf{T}}
\newcommand{\I}{\mathrm{i}}
\newcommand{\e}{\mathrm{e}}
\renewcommand{\Re}{\operatorname{Re}}

\newcommand{\<}{\left\langle}	
\renewcommand{\>}{\right\rangle}	

\newcommand{\mss}{\mathfrak{s}}
\newcommand{\mst}{\mathfrak{t}}
\newcommand{\msu}{\mathfrak{u}}
\newcommand{\msw}{\mathfrak{m}}

\newcommand{\df}[3]{\prescript{#2}{#3}{B}^{#1}}

\newcommand{\kernelF}[2]{\prescript{#1}{#2}{K}}
\newcommand{\kernelFG}[2]{\prescript{#1}{#2}{K}}

\newcommand{\dgrfSE}{\Sigma}

\newcommand{\gGN}{g_{\scriptscriptstyle{\text{GN}}}}
\newcommand{\gGNCoeff}[1]{g_{{\scriptscriptstyle \text{GN}}, #1}}
\newcommand{\gTh}{g_{\scriptscriptstyle{\text{Th}}}}
\newcommand{\gThCoeff}[1]{g_{{\scriptscriptstyle\text{Th}}, #1}}
\newcommand{\gNJL}{g_{\scriptscriptstyle{\text{NJL}}}}
\newcommand{\subGN}[1]{{#1}_{\scriptscriptstyle{\text{GN}}}}
\newcommand{\subTh}[1]{{#1}_{\scriptscriptstyle{\text{Th}}}}
\newcommand{\subNJL}[1]{{#1}_{\scriptscriptstyle{\text{NJL}}}}
\newcommand{\Nf}{\ensuremath{N_{\text f}}}
\newcommand{\Nfcrit}{\ensuremath{N_{\text{f,crit}}}}

\allowdisplaybreaks

\begin{document}

\title{Momentum dependence of quantum critical Dirac systems}

\author{Lennart Dabelow}
\email[Electronic address: ]{ldabelow@physik.uni-bielefeld.de}
\affiliation{Fakult\"at f\"ur Physik, Universit\"at Bielefeld, 33615 Bielefeld, Germany}

\author{Holger Gies}
\email[Electronic address: ]{holger.gies@uni-jena.de}
\affiliation{Theoretisch-Physikalisches Institut, Abbe Center of Photonics, Friedrich-Schiller-Universit\"at Jena, Max-Wien-Platz 1, 07743 Jena, Germany}
\affiliation{Helmholtz-Institut Jena, Fr\"obelstieg 3, 07743 Jena, Germany}

\author{Benjamin Knorr}
\email[Electronic address: ]{b.knorr@science.ru.nl}
\affiliation{Institute for Mathematics, Astrophysics and Particle Physics (IMAPP), Radboud University Nijmegen, Heyendaalseweg 135, 6525 AJ Nijmegen, The Netherlands}


\begin{abstract}
  We analyze fermionic criticality in relativistic 2+1 dimensional
  fermion systems using the functional renormalization group (FRG),
  concentrating on the Gross-Neveu (chiral Ising) and the Thirring
  model. While a variety of methods, including the FRG, appear to reach
  quantitative consensus for the critical regime of the Gross-Neveu
  model, the situation seems more diverse for the Thirring model with
  different methods yielding vastly different results. We present a
  first exploratory FRG study of such fermion systems including
  momentum-dependent couplings using pseudo-spectral methods. Our
  results corroborate the stability of results in Gross-Neveu-type
  universality classes, but indicate that momentum dependencies become
  more important in Thirring-type models for small flavor numbers. For
  larger flavor numbers, we confirm the existence of a non-Gaussian
  fixed point and thus a physical continuum limit. In the large-$N$
  limit, we obtain an analytic solution for the momentum dependence of
  the fixed-point vertex.
\end{abstract}

\maketitle

\section{Introduction}
\label{sec:Introduction}

Surprisingly many condensed-matter systems support gapless
Dirac-fermionic modes as emergent low-energy degrees of freedom,
exhibiting a quasi-relativistic dispersion relation.
This plethora of systems includes, for example, materials of graphene type, ${}^3$He, surface states of
topological insulators, or $d$-wave cuprate superconductors.
Despite being rather different microscopically, these systems are summarized as the class of Dirac materials, see e.g. Refs.~\onlinecite{Vafek:2013mpa,Wehling:2014cla} for reviews.

The corresponding investigation of low-energy effective theories for such systems and associated phase transitions, e.g., semimetal-insulator transitions, has been a focus of theoretical research in recent years \cite{Sorella_1992,PhysRevLett.87.246802, PhysRevLett.97.146401,PhysRevB.80.081405,
PhysRevB.86.121402, PhysRevB.66.045108, PhysRevB.75.134512, PhysRevLett.100.146404, Herbut:2009vu, PhysRevB.79.085116,
PhysRevLett.100.156401, Mesterhazy:2012ei,PhysRevB.87.085136, PhysRevB.89.035103, Janssen:2014gea, PhysRevLett.98.186809,
PhysRevB.82.035429, PhysRevB.85.155439, PhysRevB.90.035122, PhysRevB.92.155137, PhysRevLett.106.236805,
PhysRevLett.86.958, PhysRevB.78.165423, PhysRevB.81.125105, 1751-8121-45-38-383001, PhysRevB.81.085105, PhysRevB.87.041401,
PhysRevB.82.205106, PhysRevD.88.065008, Jakovac:2014lqa, Scherer:2016zwz, Choi:2016sxt, Mihaila:2017ble, Classen:2017hwp,  Iliesiu:2017nrv, Huffman:2017swn, Zerf:2017zqi}. These theories are considered to transcend the standard Landau-Ginzburg-Wilson paradigm for critical phenomena, as gapless fermions remain present at the phase transition \cite{Hofling:2002hj,Classen:2017hwp,PhysRevB.97.125137} possibly being subject to further spectator symmetries \cite{Gehring:2015vja}.

The present work is devoted to a study of two well-known models of relativistic Dirac fermions in 2+1 dimensional spacetime: the (irreducible) Gross-Neveu model \cite{Gross:1974jv} and the Thirring model \cite{Thirring:1958in}. Both models are classically defined in terms of a local four-fermion interaction, and exhibit a U(2$\Nf$) symmetry, where $\Nf$ denotes the number of massless (reducible 4-component) Dirac flavors. Both models are naively ``perturbatively nonrenormalizable'', but have shown to be nonperturbatively renormalizable, e.g. to any order in a $1/\Nf$ expansion \cite{Parisi:1975im,Gawedzki:1985ed,Rosenstein:1988pt,deCalan:1991km,ZinnJustin:1991yn}, and thus are simple examples \cite{Braun:2010tt} of the concept of asymptotic safety \cite{Weinberg:1976xy}. Despite their similarity, the long-range behavior can be rather different. As a function of the coupling as a control parameter, the Gross-Neveu model exhibits a second-order quantum phase transition to a gapped phase for any value of the flavor number $\Nf$. The order parameter is a scalar in flavor space, breaking a parity-like discrete $\mathbbm{Z}_2$ symmetry in the gapped phase. Correspondingly, the universality class associated with fermionic criticality in this case is also called the chiral Ising universality class \cite{Classen:2015ssa}. Quantitative studies of universal critical phenomena have been performed for more than two decades \cite{Luperini:1991sv,Rosenstein:1993zf,Gracey:1993kc,Gracey:1993kb,Hands:1992be,Karkkainen:1992sk,Karkkainen:1993ef,Vasiliev:1992wr, Focht:1995ie,PhysRevLett.86.958,Hofling:2002hj,Braun:2010tt, Chandrasekharan:2013aya, Wang:2014cbw, Li:2014aoa}. By now, a quantitative consensus for the critical exponents is beginning to emerge \cite{Vacca:2015nta,Gracey:2016mio, Fei:2016sgs, Knorr:2016sfs, Hands:2016foa, Schmidt:2016rtz, Mihaila:2017ble, Iliesiu:2017nrv, Huffman:2017swn,Zerf:2017zqi,Ihrig:2018hho}.

By contrast, the Thirring model does not exhibit a phase transition at large
$\Nf$ in agreement with results in the strict $\Nf\to\infty$ limit
\cite{PhysRevD.43.3516, Hands:1994kb}. Also, the order parameter is a
bifundamental scalar in flavor space \cite{Janssen:2012pq}, describing the
breaking to a U($\Nf$)$\otimes$U($\Nf$) subgroup in case of gap
formation. Early studies using a variety of methods suggested the existence of
a critical flavor number $\Nfcrit$, above which no gapped phase exists, and
below which a transition to a gapped phase occurs at sufficiently strong
coupling. Quantitative estimates of $\Nfcrit$ since then have spanned a wide
range of values \cite{PhysRevD.43.3516, Hong:1993qk,
  Itoh:1994cr,Kondo:1995np,Kim:1996xza, DelDebbio:1997dv, Hands:1999id,
  Christofi:2007ye, Gies:2010st,Chandrasekharan:2011mn}. As analyzed in
Ref.~\onlinecite{Janssen:2012pq}, various reasons may have led to this
diversity, e.g., also the presence/absence of exact chiral symmetry in the
different methods. Recent lattice simulations with exact chiral symmetry
\cite{Hands:2016foa,Hands:2018vrd, Wellegehausen:2017goy,Lenz:2019qwu} indicate
that there may be no symmetry breaking for any integer $\Nf\geq 2$ at
all. Whereas simulations using domain wall fermions provide evidence for
$0<\Nfcrit<2$ with some indications for $\Nfcrit>1$ \cite{Hands:2018vrd}, the
latest results of Ref.~\onlinecite{Lenz:2019qwu} using SLAC fermions and an analytic continuation of the
parity-even theory to arbitrary real $\Nf$ show clear evidence that the
critical flavor is below unity, $\Nfcrit<1$.
 
In view of these similarities and differences between the two models,  we wish to address three immediate questions in this work: \textit{(i)} Is there a way to quantify the higher degree of complexity of the Thirring model compared with the Gross-Neveu model? \textit{(ii)} If so, can this information be used to construct better approximation schemes for the Thirring model? \textit{(iii)} Is the Thirring model a UV complete (asymptotically safe) quantum field theory, despite the absence of a phase transition for a wide range of $\Nf$ values? While the second question aims at an improvement of analytical methods, the answer to the third question is also relevant for the existence of a continuum limit of discrete lattice realizations.

For addressing these questions, we need a theoretical tool that can simultaneously deal with both models in the continuum, preferably for all values of $\Nf$ and directly in $d=2+1$ dimensions. These criteria are met by the functional renormalization group (\FRG{}) which offers a substantial degree of flexibility of devising systematic nonperturbative approximation schemes. In fact, the \FRG{} as formulated in terms of the Wetterich equation \cite{Wetterich:1992yh} has already widely been used for the two models under consideration. Common approximation schemes focus on local fermionic interactions \cite{Gies:2010st,PhysRevD.88.065008,Jakovac:2014lqa,Gehring:2015vja}, or use partial-bosonization techniques together with a derivative expansion to also account for the emerging composite bosonic degrees of freedom \cite{PhysRevLett.86.958,Hofling:2002hj,Braun:2010tt,Janssen:2012pq,Mesterhazy:2012ei,Janssen_2012,Janssen:2014gea,Vacca:2015nta,Borchardt:2015rxa,Knorr:2016sfs,Knorr:2017yze}. Whereas these methods work very well for the Gross-Neveu model, exhibiting apparent convergence for the quantitative estimates of critical exponents \cite{Knorr:2016sfs}, the Thirring model already on this level is more involved, requiring composite scalar and vector boson fields \cite{Janssen:2012pq} as well as dynamical bosonization techniques \cite{Gies:2001nw,Pawlowski:2005xe,Floerchinger:2009uf}.

In the present work, we go beyond the limitations of these \FRG{} approximation schemes by investigating the momentum dependence of the fermionic self-interactions. In this manner, we can specifically address the question as to whether the assumptions underlying the use of the derivative expansion are justified. For the present exploratory study, we concentrate specifically on the momentum dependence arising in the $\mss{}$ channel, using modern pseudo-spectral methods as developed in Refs.~\onlinecite{Borchardt:2015rxa, Borchardt:2016pif} to resolve this dependence to a high accuracy.

Our results provide at least partial affirmative answers to all three questions raised above: \textit{(i)} The quantitative relevance of momentum dependence provides for a measure to assess the higher degree of complexity of the Thirring model. By contrast, our results for the Gross-Neveu model lend strong support to the quantitative use of the derivative expansion. \textit{(ii)} More adapted approximation schemes for the Thirring model hence have to resolve the more complex momentum dependence of the fermionic self-interactions in particular for smaller flavor numbers; for future work, a combination with dynamical bosonization techniques absorbing momentum dependencies in the composite channels appears recommended. \textit{(iii)} Our results provide further evidence for the existence of a non-Gaussian Thirring fixed point for a wide range of flavor numbers. The 2+1 dimensional Thirring model hence can be a UV-complete quantum field theory, also implying the existence of a continuum limit of lattice formulations independently of the observation of a specific chiral-symmetry breaking phase transition.

This work is structured as follows: in Section~\ref{sec:FRG} we shortly discuss the relevant aspects of the \FRG{}. Section~\ref{sec:MomDepCouplings} discusses the general structure of the renormalization group equations, whereas in Section~\ref{sec:GNThFlowEquations} we present the explicit flow equations. 
After reviewing the complementary bosonization approach and considering some limiting cases in Sections~\ref{sec:PartialBosonization} and~\ref{sec:LimitingCases}, we solve the flow equations for particular values of the flavor number in Section~\ref{sec:FiniteNf}.
We finish with a discussion of the results and a conclusion in Section~\ref{sec:Conclusions}.

\section{Functional renormalization group for fermion systems}
\label{sec:FRG}

We consider theories with $\Nf$ flavors of relativistic fermions described by the Grassmann-valued spinor fields $\psi_i$, 
$\bar\psi^i$ ($i = 1, \ldots, \Nf$) in $d=3$ Euclidean dimensions.
We also define a collective field variable $\Psi := (\bar\psi, \psi^\T)^\T$ in the spirit of Nambu-Gorkov spinors.
For the spin base, we use $\gamma$ matrices, $\gamma_\mu \in \mathbb{C}^{d_\gamma \times d_\gamma}$, generating the Clifford algebra $\lbrace \gamma_\mu, 
\gamma_\nu \rbrace = 2 \delta_{\mu\nu}$. We begin with a reducible representation with $d_\gamma = 4$. A transition to an irreducible representation is described below. 

For the following nonperturbative investigation, we use the functional renormalization group (\FRG) which facilitates to systematically integrate out quantum fluctuations at all energy scales. We specifically make use of the \FRG\ to compute the effective 
action $\Gamma$ of a theory starting from its classical, microscopic action $S$. The key ingredient is a scale-dependent effective average action $\Gamma_k$ that only 
includes fluctuations with momenta $p^2 \gtrsim k^2$. Thereby, $\Gamma_k$ interpolates between the microscopic action $\Gamma_\Lambda \simeq S$ for some 
high-energy scale $k=\Lambda$ and the full effective action $\Gamma_{k=0} = \Gamma$. For this, we add a scale-dependent regulator term in momentum space to the classical action, 
\begin{equation}
\label{eq:DefRegulator}
	\Delta S_k[\Psi] := \frac{1}{2} \int_p \Psi(-p)^\T \mc R_k(p) \Psi(p),
\end{equation}
which effectively decouples modes with momenta $p^2 \lesssim k^2$ from the theory. We use symmetry preserving regulators of the form $\mc R_k(p)=-\slashed{p}\, r(p^2/k^2)$ with a dimensionless regulator shape function $r$ that diverges sufficiently fast for small argument to provide for a $k$-dependent gap of the fermionic propagator at low momenta. Note that $\int_p \equiv \int \frac{\d^d 
p}{(2\pi)^d}$ abbreviates a normalized momentum integral. For generality, we keep the spacetime dimension formally arbitrary here and in the following; however, the Dirac structure considered explicitly in this work is tied to the vicinity of $d=3$ dimensions. The average action is then defined as a modified Legendre transform of the generating functional of connected 
correlators,
see Refs.~\onlinecite{Berges:2000ew, Pawlowski:2005xe, Gies:2006wv, Delamotte:2007pf, 
Kopietz:2010zz, Rosten:2010vm, Braun:2011pp, Sfondrini:2012wu,RevModPhys.84.299} for reviews. A cornerstone of the \FRG\ formalism is given by the Wetterich 
equation \cite{Wetterich:1992yh, Bonini:1992vh, Bornholdt:1992za, Ellwanger:1993mw, Morris:1993qb}
\begin{align}
\label{eq:RG:WetterichEquation}
    \partial_t \Gamma_k = \frac{1}{2} \STr \left[ \left( \Gamma_k^{(2)} + \mc R_k \right)^{-1} \partial_t \mc R_k \right] , \quad t = \ln\left( 
\frac{k}{\Lambda} \right) ,
\end{align}
representing an exact functional differential equation for $\Gamma_k$ that describes the flow from the microscopic action to the full effective action. Here,  
$\Gamma_k^{(2)}$ denotes the second functional derivative with respect to the fields $\Psi$, and the supertrace STr is a shorthand for a summation over discrete and integration over continuous variables. 

While \Eqref{eq:RG:WetterichEquation} is an exact equation, exact solutions are generally difficult to obtain. In practice, it is useful to work with systematic truncation schemes, concentrating on an accessible subset of theory space.
There are two common choices to truncate the theory space. A derivative expansion spans the action in terms of operators sorted according to the number of derivatives acting on the quantum fields; the advantage is that full field dependencies can be retained. This approximation scheme is especially useful for investigating the action of bosonic order-parameter fields in the study of spontaneous symmetry breaking and the description of condensation effects. By contrast, a vertex expansion spans the action in terms of correlation functions with an increasing number of external legs. This allows to address the full momentum dependence of the correlations, having become a successful strategy to explore gauge theories and gravity where the couplings are inherently momentum dependent. 

For fermionic systems as considered here, previous studies have concentrated on either lowest-order derivative expansions in a purely fermionic description \cite{Gies:2010st,Jakovac:2014lqa,Gehring:2015vja}, or derivative expansions in a partially bosonized description within the so-called LPA' approximation \cite{PhysRevLett.86.958,Hofling:2002hj,Braun:2010tt,Janssen:2012pq,Mesterhazy:2012ei,Janssen:2014gea}, including multi-bosonic interactions \cite{Vacca:2015nta}; for some models, even complete next-to-leading order results are available \cite{Knorr:2016sfs,Knorr:2017yze}.

In this work, we use a vertex expansion to study the momentum
dependence of four-fermion couplings for the first time for the
present set of models. More precisely, we account for the full
momentum dependence of the propagator, a projected choice of momentum
dependencies of the four-fermion vertices, and neglect possible
contributions from higher vertices.

Resolving momentum dependent propagators and the flow of higher-order vertex functions has become a central tool for analyzing flow equations for the theory of the strong interactions \cite{Meggiolaro:2000kp,Ellwanger:1995qf,Cyrol:2016tym,Cyrol:2017ewj} as well as quantum gravity \cite{Christiansen:2014raa,Christiansen:2015rva,Denz:2016qks}. Within the FRG, also approximation strategies to resolve both momentum and field dependence are available \cite{Benitez:2009xg, Benitez:2011xx}. The present work is among the few examples, where not only the full propagator, but also the functional dependence of vertex functions on a momentum channel are resolved near an interacting fixed point.

\section{Flow of momentum-dependent couplings}
\label{sec:MomDepCouplings}

\begin{figure*}[tb]
\begin{minipage}{\textwidth}
\begin{minipage}{0.32\textwidth}
\flushleft\textbf{(a)}\ \\
\centering
	\includegraphics{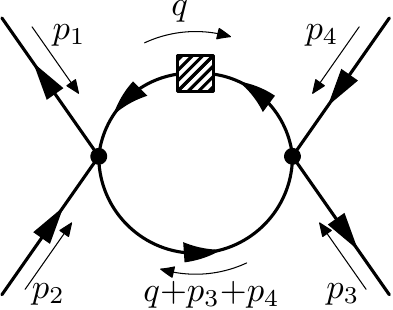}
\end{minipage}		
\begin{minipage}{0.32\textwidth}
\flushleft\textbf{(b)}\ \\
\centering
	\includegraphics{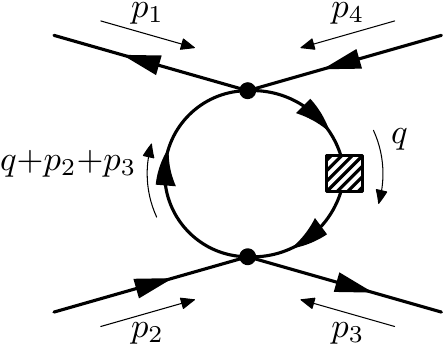}
\end{minipage}
\begin{minipage}{0.32\textwidth}
\flushleft\textbf{(c)}\ \\
\centering
	\includegraphics{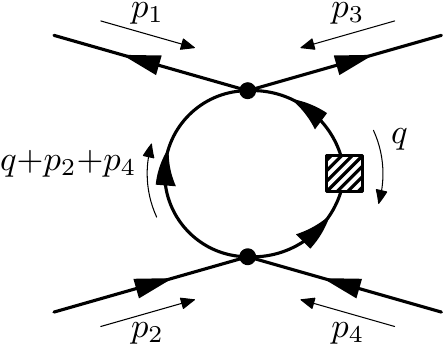}
\end{minipage}
\end{minipage}
\caption{Feynman diagrams for the different Mandelstam channels of regularized one-loop processes contributing to the \RG{} flow of a generic four-fermion 
interaction vertex. The shaded boxes denote the regulator insertions. The three channels are conveniently parametrized in terms of the Mandelstam variables, 
which encode the momentum transferred across the loop. Diagram (a) corresponds to the $\mss$ channel, diagram (b) to the $\mst$ channel, and diagram (c) to 
the $\msu$ channel.}
\label{fig:FourFermionInteraction:RGFlowDiagrams}
\end{figure*}

We consider models defined on the basis of the flavor and Dirac structure of four-fermion interactions, \ie\ models with local interactions of the form
\begin{equation}
\label{eq:FourFermionVertex}
\begin{aligned}
	&\int \d^d x \; \bar g \, \bar\psi(x) \mc O_1 \psi(x) \, \bar\psi(x) \mc O_2 \psi(x) \\
	&\quad = \int_{p_2,p_3,p_4} \!\!\!\!\!\!\!\!  \bar g \, \bar\psi(-p_1) \mc O_1 \psi(p_2) \, \bar\psi(-p_3) \mc O_2 \psi(p_4) \,,
\end{aligned}
\end{equation}
where $p_1 \equiv -p_2-p_3-p_4$ in the Fourier transformed version in
the second line. Our convention assumes all momenta as in-coming into
the vertex. The coupling $\bar g$ parametrizes the strength of the
interaction and $\mc O_1$ and $\mc O_2$ are flavor and Dirac space
matrices characterizing the vertex. In the following, we mostly
consider the Gross-Neveu \cite{Gross:1974jv} and Thirring
\cite{Thirring:1958in} models defined in terms of $\subGN{\mc O} = \bm
1 \otimes \bm 1$ and $\subTh{\mc O} = \bm 1 \otimes \gamma_\mu$,
respectively, for both $\mc O_1$ and $\mc O_2$.  Here, the first
factor of the tensorial structure refers to the flavor index and the
second to Dirac space.

Upon integrating out fluctuations, the corresponding 4-point
correlators will acquire nonlocal contributions. These can be
parametrized by promoting the coupling in~\eqref{eq:FourFermionVertex}
to a function of the external momenta, $\bar g = \bar g(p_1, p_2, p_3,
p_4)$. Because of momentum conservation, only three of these vectors
are independent. Lorentz symmetry reduces the number of variables to a
maximum of six linearly independent invariants that can be constructed
from three independent momentum vectors. It is still useful to stick
to the redundant notation above, as it allows us to keep
track of the flow of loop momenta in the Wetterich
equation~\eqref{eq:RG:WetterichEquation} and goes along well with the
pictorial language of Feynman diagrams.

Since we are eventually interested in fixed points and critical
behavior of specific models, we switch to dimensionless quantities by
measuring all dimensionful quantities in units of the RG scale
$k$. \text{E.g.} we use dimensionless momenta $\tilde p = p / k$, and
also introduce renormalized fields $\tilde{\psi}(\tilde p) :=
Z^{1/2}(p^2) \psi(p)$, $\tilde{\bar\psi}(p) := Z^{1/2}(p^2) \psi(p)$,
where $Z(p^2)$ denotes a momentum-dependent wave function
renormalization. Similarly, the dimensionless, renormalized
four-fermion couplings parametrizing the 4-point correlator are
defined as $\tilde g(\tilde p_i) := k^{d-2} \left[ Z(p_1^2) Z(p_2^2)
  Z(p_3^2) Z(p_4^2) \right]^{-1/2} \bar g(p_i)$. This rescaling leads
to additional contributions to the flow equation of the couplings,
\begin{equation}
\label{eq:MomDepFourFermion:FlowEquation:ScalingTerm}
\begin{aligned}
	\partial_t \tilde g(\tilde p_i) &= \left[ d - 2 + \frac{1}{2} \sum\nolimits_{j=1}^4 \eta(\tilde p_j^2) \right] \tilde g(\tilde p_i) \\
		&\qquad + \sum_{j=1}^4 \tilde p_j \cdot \nabla_{\tilde p_j} \tilde g(\tilde p_i) + \ldots,
\end{aligned}
\end{equation}
where the ellipsis stands for the quantum fluctuations, \ie\ the
loop-induced flow of $\bar g$. Here, we have also introduced a
generalized, momentum-dependent anomalous dimension $\eta(p^2) :=
-\partial_t \ln Z(p^2)$.  In the following, the dimensionless notation
is implicitly understood, so that we drop all tildes and use the bar
notation to distinguish dimensionful couplings from their
dimensionless counterparts.

The Wetterich equation~\eqref{eq:RG:WetterichEquation} has a one-loop structure.
A general analysis of generic four-fermion vertices shows that there are four different processes or diagrams contributing to the \RG\ flow of 
the coupling functions. These are depicted in Fig.~\ref{fig:FourFermionInteraction:RGFlowDiagrams}, where the process in panel (a) exists in two 
variants, see below. The diagrams are characterized by different momentum transfers across the loop and are conveniently labeled in terms of the Mandelstam variables 
\cite{Mandelstam:1958xc} $\mss = (p_1+p_2)^2 = (p_3+p_4)^2$, $\mst = (p_1+p_4)^2 = (p_2+p_3)^2$, and $\msu = (p_1+p_3)^2 = (p_2+p_4)^2$. In this spirit, we 
refer to the diagrams in Fig.~\ref{fig:FourFermionInteraction:RGFlowDiagrams} (a)-(c) as $\mss$, $\mst$, and $\msu$ channel processes. We emphasize that the diagrams denote processes involving full propagators and fully-momentum dependent vertices and hence are nonperturbative.

\begin{figure}[tb]
\phantom{x}

\phantom{x}
    \includegraphics{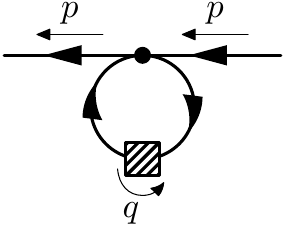}
\caption{Self-energy diagram encoding the contributions from the four-fermion coupling that enter the anomalous dimension function.}
\label{fig:FourFermionInteraction:SelfEnergyDiagram}
\end{figure}

In order to cast them into a more explicit form, we introduce the \emph{threshold kernel}
\begin{equation}
\label{eq:ThresholdKernel}
\begin{aligned}
	\kernelF{a_1}{a_2}(p, q; \eta)
		&:= \left[ \frac{ (a_1 + a_2) q^2 + a_1 (q \cdot p) - a_2 \frac{(q \cdot p)^2}{p^2} }{ q^2 (q+p)^2} \right] \\
		&\qquad \times \frac{ \partial_t r(q^2) - \eta(q^2) r(q^2) }{ \left[ 1+r(q^2) \right]^2 \left[ 1 + r([q+p]^2) \right] },
\end{aligned}
\end{equation}
which encodes the propagation of the loop fermions. Here $q$ denotes the loop momentum, $p$ the momentum transfer, $\eta$ is the anomalous dimension function, and the regulator shape function $r$ was introduced below Eq.~\eqref{eq:DefRegulator}. 
By convention, we always choose $q$ to be the momentum flowing through the regulator insertion. The prefactors $a_1$ and $a_2$ parametrize different momentum 
modes of the propagating loop fermion. The threshold kernel is related to the threshold function $\ell^{(F)}_1$ familiar from RG flows in the pointlike limit (see, \eg, 
Ref.~\onlinecite{Janssen:2012pq}) via the relation
\begin{equation}
	\int_q \kernelF{a_1}{a_2}(0, q; \eta) = 2 v_d \left( a_1 + \frac{d-1}{d} a_2 \right) \ell^{(F)}_1(0; \eta)
	\label{eq:ThresholdKernel:RelationToThresholdFunction}
\end{equation}
with $v_d = \left[ 2^{d+1} \pi^{d/2} \Gamma \left( \frac{d}{2} \right) \right]^{-1}$. The diagrams in 
Fig.~\ref{fig:FourFermionInteraction:RGFlowDiagrams} then are in correspondence with the \emph{diagram functionals}
\begin{widetext}
\begin{subequations}
\label{eq:DiagramFunctionals}
\begin{align}
	\df{\mss 2}{a_1}{a_2}[g_1, g_2; \eta](p_i) &:= \frac{1}{\tilde N} \int_q \kernelF{a_1}{a_2}(p_3\!+\!p_4, q; \eta) \left[ g_1(p_1, p_2, q^\prime, -\!q) 
g_2(q, -\!q^\prime, p_3, p_4) + \, (q \longleftrightarrow -\!q^\prime) \vphantom{A^A_A}\right]_{q^\prime = q\!+\!p_3\!+\!p_4}, \displaybreak[0] \\
	\df{\mss 1}{a_1}{a_2}[g_1, g_2; \eta](p_i) &:= \frac{1}{\tilde N} \int_q \kernelF{a_1}{a_2}(p_3\!+\!p_4, q; \eta) \left[ g_1(p_1, p_2, q^\prime, -\!q) 
g_2(p_3, -\!q^\prime, q, p_4) + \, (q \longleftrightarrow -\!q^\prime) \vphantom{A^A_A}\right]_{q^\prime = q\!+\!p_3\!+\!p_4}, \displaybreak[0] \\
	\df{\mst}{a_1}{a_2}[g_1, g_2; \eta](p_i) &:= \frac{1}{\tilde N} \int_q \kernelF{a_1}{a_2}(p_2\!+\!p_3, q; \eta) \left[ g_1(p_1, q^\prime, -\!q, p_4) 
g_2(-\!q^\prime, p_2, p_3, q) + \, (q \longleftrightarrow -\!q^\prime) \vphantom{A^A_A}\right]_{q^\prime = q\!+\!p_2\!+\!p_3}, \displaybreak[0] \\
	\df{\msu}{a_1}{a_2}[g_1, g_2; \eta](p_i) &:= \frac{1}{\tilde N} \int_q \kernelF{a_1}{a_2}(p_2\!+\!p_4, q; \eta) \left[ g_1(p_1, q^\prime, p_3, -\!q) 
g_2(-\!q^\prime, p_2, q, p_4) + \, (q \longleftrightarrow -\!q^\prime) \vphantom{A^A_A}\right]_{q^\prime = q\!+\!p_2\!+\!p_4},
\end{align}
\end{subequations}
\end{widetext}
where $\tilde N = \Nf d_\gamma = 4 \Nf$ counts the independent spinor components.
The notation ``$+\,(q \longleftrightarrow -q^\prime)$'' here implies that all preceding terms within the same bracket level have to be added with $q$ and $-q^\prime$ 
interchanged. In relation to the diagrams, we identify the left/top vertices with the coupling $g_1$ and the right/bottom vertices with the coupling $g_2$. The two distinct versions of the $\mss$-channel diagram mentioned above differ by an exchange of the $1$- and $3$-legs in the second vertex.

In a similar fashion, we can analyze the self-energy contributions from the four-fermion couplings entering the flow of the wave function renormalization. The 
corresponding diagram, Fig.~\ref{fig:FourFermionInteraction:SelfEnergyDiagram}, exists in two variants again. The associated diagram functionals are
\begin{widetext}
\begin{subequations}
\label{eq:DiagramFunctionals:SelfEnergy}
\begin{align}
	\dgrfSE^1[g; \eta](p^2) &:= -\frac{1}{\tilde N} \int_q \left( \frac{ q \cdot p }{ q^2 p^2 } \right) \frac{\partial_t r(q^2) - \eta(q^2) 
r(q^2)}{\left[ 1+r(q^2) \right]^2} g(-p, -q, q, p) \,, \displaybreak[0] \\
	\dgrfSE^{2}[g; \eta](p^2) &:= -\frac{1}{\tilde N} \int_q \left( \frac{ q \cdot p }{ q^2 p^2 } \right)  \frac{ \partial_t r(q^2) - \eta(q^2) r(q^2) 
}{ \left[ 1+r(q^2) \right]^2 } g(-p, p, -q, q) \,.
\end{align}
\end{subequations}
\end{widetext}
These definitions will allow us to express the flows of wave function renormalizations and coupling functions of arbitrary four-fermion models to fourth order in the vertex expansion.

\section{Flow equations of the Gross-Neveu-Thirring model}
\label{sec:GNThFlowEquations}

After these general considerations, we now study a class of models defined
by the symmetries of the Thirring model. The Thirring
model is defined by the microscopic action \cite{Thirring:1958in}
\begin{equation}
\label{eq:ThModel:MicroscopicAction}
	S[\psi, \bar\psi] = \int \d^d x \left[ \bar\psi(x) \I \slashed\partial \psi(x) + \frac{\subTh{\bar g}}{2 \tilde N} \left( \bar\psi(x) \gamma_\mu 
\psi(x) \right)^2  \right] ,
\end{equation}
with the convention $\tilde N = \Nf d_\gamma = 4 \Nf$, as introduced below \Eqref{eq:DiagramFunctionals}.
The vector-type Thirring interaction $(\bar\psi \gamma_\mu \psi)^2$
exhibits obvious invariances under $\mathrm U(\Nf)$ flavor rotations
as well as continuous transformations $\psi \mapsto \e^{\I \alpha
  M_\pm} \psi$, $\bar\psi \mapsto \bar\psi \e^{\pm \I \alpha M_{\pm}}$
generated by the Dirac algebra elements $M_+ \in \lbrace \gamma_4,
\gamma_5 \rbrace$ and $M_- \in \lbrace \bm 1, \I \gamma_{4} \gamma_{5}
\rbrace$ (forming a $\mathrm U(2)$ group). Because of the partial
similarity to chiral or axial transformations in $d=4$, the latter
symmetry is often called ``chiral'' in the literature. In fact, these
two symmetries are actually subgroups of a larger $\mathrm U(2\Nf)$
symmetry, which will become more obvious in the irreducible spinor
formulation introduced below. In addition, the model also exhibits
discrete $C$, $P$, and $T$ symmetries which can be formulated in
various ways \cite{Janssen:2012pq}. Up to Fierz transformations, there
is only one more linearly independent local four-fermion interaction
sharing the symmetries of the Thirring model \cite{Gies:2010st,
  Janssen_2012, Janssen:2012pq, Gehring:2015vja}. In our conventions,
this can be written as $(\bar\psi \gamma_{45} \psi)^2$ with
$\gamma_{45} := \I \gamma_4 \gamma_5$.

The $\mathrm U(2\Nf)$ symmetry of the model becomes explicit by
decomposing the $\Nf$ four-component Dirac spinors $\psi$, $\bar\psi$
into $2\Nf$ two-component Weyl spinors $\chi$, $\bar\chi$ transforming
in the irreducible representation of the Clifford algebra
\cite{Gies:2010st, Janssen_2012, Janssen:2012pq}. Extracting $\chi$,
$\bar\chi$ as chiral projections of $\psi$, $\bar\psi$ by means of the
projectors $P_\pm = \frac{1}{2} (\bm 1 \pm \gamma_{45})$, the fermion
bilinears are mapped into
\begin{equation}
	\bar\psi \gamma_{45} \psi \mapsto \bar\chi \chi
	\quad \text{ and } \quad
	\bar\psi \gamma_\mu \psi \mapsto \bar\chi \sigma_\mu \chi
\end{equation}
with $\sigma_\mu$ denoting the Pauli matrices,
\begin{equation}
\label{eq:PauliMatrices}
	\sigma_1 = \lmat 0 & 1 \\ 1 & 0 \rmat ,
	\quad
	\sigma_2 = \lmat 0 & -\I \\ \I & 0 \rmat ,
	\quad
	\sigma_3 = \lmat 1 & 0 \\ 0 & -1 \rmat .	
\end{equation}
The interactions discussed above hence correspond to the Thirring
interaction and the Gross-Neveu interaction \cite{Gross:1974jv} of the irreducible
representation, featuring an explicit $\mathrm U(2\Nf)$ symmetry.

Even if the bare action was defined by the Thirring interaction \eqref{eq:ThModel:MicroscopicAction} at some high scale, fluctuations would generically generate also the Gross-Neveu interactions. Therefore, we include both interaction structures in our ansatz for the effective average action, which reads in 
momentum space:
\begin{equation}
\begin{aligned}
	&\Gamma_k[\chi, \bar\chi] = - \int_p Z(p^2) \bar\chi(p) \slashed p \chi(p) \\
		&\quad + \int_{p_2,p_3,p_4} \!\!\!\! \frac{\subGN{\bar g}(p_i)}{2 \tilde N} \bar\chi(-p_1) \chi(p_2) \bar\chi(-p_3) \chi(p_4) \\
		&\quad + \int_{p_2,p_3,p_4} \!\!\!\! \frac{\subTh{\bar g}(p_i)}{2 \tilde N} \bar\chi(-p_1) \sigma_\mu \chi(p_2) \bar\chi(-p_3) \sigma^\mu 
\psi(p_4).
\end{aligned}
	\label{eq:GNTh:EffectiveAverageAction}
\end{equation}
Inserting $\Gamma_k$ into the Wetterich equation~\eqref{eq:RG:WetterichEquation} and ignoring contributions from $n>4$-point correlators, we can extract the flow equations 
or $\beta$ functions for the dimensionless couplings,
\begin{widetext}
\begin{subequations}
\label{eq:GNTh:BetaFunctions}
\begin{align}
\label{eq:GNTh:BetaFunction:G}
&\begin{aligned}
	\partial_t \gGN &= \left[ 1 + 2 \eta^{(4)} + \sum\nolimits_{j} p_j \cdot \nabla_{p_j}\right] \gGN - \df{\mss 2}{\tilde N}{0}[\gGN, \gGN] + \df{\mss 
1}{2}{0}[\gGN, \gGN] + \df{\mss 1}{3}{0}_\leftrightarrow[\gGN, \gTh] \\
		&\qquad + \df{\mst}{1}{0}_\leftrightarrow[\gGN, \gTh] + \df{\mst}{2}{0}[\gTh, \gTh]  - \df{\msu}{1}{0}_\leftrightarrow[\gGN, \gTh] + 
\df{\msu}{2}{0}[\gTh, \gTh] \,, \\
\end{aligned} \\
\label{eq:GNTh:BetaFunction:V}
&\begin{aligned}
	\partial_t \gTh &= \left[ 1 + 2 \eta^{(4)} + \sum\nolimits_{j} p_j \cdot \nabla_{p_j}\right] \gTh \quad+ \df{\mss 2}{\tilde N}{-\tilde N}[\gTh, \gTh] - 
\df{\mss 1}{1}{-1}_\leftrightarrow[\gGN, \gTh] + \df{\mss 1}{2}{-2}[\gTh, \gTh] \\
		&\qquad + \df{\mst}{0}{1/2}[\gGN, \gGN] + \df{\mst}{1}{-1/2}_\leftrightarrow[\gGN, \gTh] + \df{\mst}{2}{1/2}[\gTh, \gTh] \\
		&\qquad - \df{\msu}{0}{1/2}[\gGN, \gGN] + \df{\msu}{1}{-1/2}_\leftrightarrow[\gGN, \gTh] - \df{\msu}{2}{1/2}[\gTh, \gTh] \,.
\end{aligned}
\end{align}
\end{subequations}
\end{widetext}
The notation $\df{\msw}{a_1}{a_2}_\leftrightarrow[g_1, g_2] \equiv \df{\msw}{a_1}{a_2}[g_1, g_2] + \df{\msw}{a_1}{a_2}[g_2, g_1]$ stands for a symmetrized 
version of the diagram functionals, and $\eta^{(4)}(p_1^2, \ldots, p_4^2) := \frac{1}{4} \sum_{j=1}^4 \eta(p_j^2)$. In addition, we suppressed the dependence on 
the anomalous dimension function $\eta$ in the diagram functionals. We observe that all four types of diagrams worked out in Sect.~\ref{sec:MomDepCouplings} 
enter the flow equations.

However, we expect that the $\mss$-channel dominates for several reasons: For the flow of $\gGN$, there are more $\mss$-channel diagrams than 
$\mst$- or $\msu$-channel ones. Also, the functions $\df{\mst}{1}{0}_\leftrightarrow[\gGN, \gTh]$ and $\df{\msu}{1}{0}_\leftrightarrow[\gGN, \gTh]$ come with 
opposite signs, such that mutual cancellations can occur. For the flow of $\gTh$, this is the case for the diagrams of order $\gGN^2$ as well as the 
ones of order $\gTh^2$. Moreover, the $\mss 2$-diagrams are the only ones remaining in the limit $\Nf \to \infty$. Our expectation of $\mss$-channel dominance is self-consistently verified below.

The flow of the wave function renormalization is governed by the anomalous dimension function which can similarly be extracted from the Wetterich equation,
\begin{equation}
\label{eq:GNTh:ADF}
	\eta(p^2) = \dgrfSE^1[\gGN; \eta] - \dgrfSE^1[\gTh; \eta] + \tilde N \dgrfSE^{2}[\gTh; \eta] .
\end{equation}
Eqs.~\eqref{eq:GNTh:BetaFunctions} and~\eqref{eq:GNTh:ADF} form a system of coupled integro-differential equations for the coupling and anomalous dimension 
functions to which, unfortunately, no exact solution is known to us.
In view of the expected $\mss$-channel dominance, we can reduce the complexity substantially, if we assume the coupling functions to depend on $\mss$ exclusively, $g(p_i) = g(\mss = [p_3+p_4]^2)$ for both $\gGN$ and $\gTh$. This \emph{$\mss$-channel approximation} 
allows us to compute fixed points and critical exponents in Sect.~\ref{sec:FiniteNf}.

\section{Partial bosonization}
\label{sec:PartialBosonization}

Let us first establish the connection between our momentum-dependent fermionic formulation and partial bosonization techniques used in previous studies \cite{PhysRevLett.86.958, Hofling:2002hj, Braun:2010tt, Braun:2011pp, Janssen:2014gea,Vacca:2015nta, Knorr:2016sfs}. Partial bosonization, in fact, allows one to recover parts of the momentum dependence of four-fermion vertices in a setting with pointlike Yukawa interactions \cite{Ellwanger:1994wy,Gies:2001nw,Braun:2011pp}. Generalizations beyond the pointlike limit have recently been suggested and studied in \onlinecite{Jakovac:2019zzw} as motivated by bound-state formation.
Moreover, partial bosonization gives direct access to the chiral condensate $\< \bar\psi \psi \>$, making it particularly useful to address questions of dynamical symmetry breaking \cite{Janssen:2012pq}.
Along the lines of a Hubbard-Stratonovich transformation \cite{Stratonovich:1957oam, Hubbard:1959ub}, we rewrite the Gross-Neveu-Thirring action~\eqref{eq:GNTh:EffectiveAverageAction} with the aid of scalar and vector boson fields $\phi$ and $V_\mu$. An ansatz for the action with local interactions in coordinate space reads in momentum space:
\begin{widetext}
\begin{equation}
\label{eq:PB:EffectiveAverageAction}
\begin{aligned}
	\Gamma_k[\phi, V, \chi, \bar\chi]
        &= \int_p \left[ - Z_{\chi} \bar\chi(p) \slashed p \chi(p) + \frac{1}{2} \phi(-p) \left( Z_{\phi} p^2 + \bar m_\phi^2 \right) \phi(p) + \frac{1}{2} \left( Z_V p^2 + \bar m_V^2 \right) V_\mu(-p) V_\mu(p) \right. \\
        		& \left. + \frac{1}{2} \left( \bar A_V - Z_V \right) p_\mu p_\nu V_\mu(-p) V_\nu(p)  \right]  + \int_{p, p^\prime} \left[ \I \bar h_\phi \phi(-p-p^\prime)  \bar\chi(-p) \chi(p^\prime) - \bar h_V V_\mu(-p-p^\prime) \bar\chi(-p) \sigma_\mu \chi(p^\prime) \right].
\end{aligned}
\end{equation}
\end{widetext}
Indeed, for vanishing kinetic terms of the boson fields ($Z_\phi = Z_V = \bar A_V = 0$), $\phi$ and $V_\mu$ become purely auxiliary Hubbard-Stratonovich fields: they can be integrated out immediately, and give back the pointlike limit of Eq.~\eqref{eq:GNTh:EffectiveAverageAction} upon identifying $\subGN{\bar g} = \tilde N \bar h_\phi^2 / \bar m_\phi^2$ and $\subTh{\bar g} = -\tilde N \bar h_V^2 / \bar m_V^2$.

As the kinetic terms are generated inevitably by the \RG\ flow, the
action \eqref{eq:PB:EffectiveAverageAction} describes momentum
transfer between pairs of fermions as mediated by the bosons. Thus,
the mixed boson-fermion theory parametrizes an effective
momentum-dependent four-fermion interaction.  In the above form, our
ansatz for the bosonized effective action corresponds to an improved
local potential approximation (\LPA'), where only the lowest order of
the effective potential, namely the mass terms $\bar m_\phi^2 \phi^2 /
2$ and $\bar m_V^2 V^2 / 2$, is kept.  Of course, it is
straightforward to include higher truncations of the effective
potential \cite{Braun:2010tt, Janssen:2012pq} or higher-order
derivatives \cite{Knorr:2016sfs}, but the present simple form suffices
for the following argument.

In order to assess the momentum dependence encoded in the bosonized action, we can map the action \eqref{eq:PB:EffectiveAverageAction} back onto a purely fermionic action. As \Eqref{eq:PB:EffectiveAverageAction} is quadratic in the boson fields, the use of the equations of motion for the boson fields is exact also on the level of the functional integral. 
Solving the classical equations of motion $\delta\Gamma_k / \delta \phi = 0$ and $\delta\Gamma_k / \delta V_\mu = 0$, we find
\begin{subequations}
\label{eq:PB:BosonVEVs}
\begin{align}
	\phi(-q) &= - \frac{ \I \bar h_\phi }{ Z_{\phi} q^2 + \bar m_\phi^2 } \int_p \bar\chi(q + p) \chi(p) \,, \\
	V_\mu(-q) &= \frac{\bar h_V}{Z_V q^2 + \bar m_V^2} \int_p \left[\vphantom{\frac{Z_V - \bar A_V}{\bar A_V q^2 + \bar m_V^2}} \bar\chi(q+p) \sigma_\mu \chi(p) \right. \notag \\
		&\quad\qquad\qquad \left. + \frac{Z_V - \bar A_V}{\bar A_V q^2 + \bar m_V^2} q_\mu \, \bar\chi(q+p) \slashed q \chi(p) \right] .
\end{align}
\end{subequations}
The first line illustrates, for instance, that the vacuum expectation value of $\phi$ is linked to the condensate $\< \bar\chi \chi \> = \< \bar\psi \gamma_{45} \psi \>$.

Substituting Eqs.~\eqref{eq:PB:BosonVEVs} back into the action~\eqref{eq:PB:EffectiveAverageAction}, we obtain the corresponding purely fermionic action,
\begin{widetext}
\begin{equation}
\begin{aligned}
	\Gamma_k
        &\simeq - \int_p Z_{\chi} \bar\chi(p) \slashed p \chi(p) 
        		+ \frac{1}{2} \int_{p_2,p_3,p_4} \!\!\!\!\!\! \frac{\bar h_V^2 (\bar A_V - Z_V) (1 + \bar A_V - Z_V)}{\left[\bar A_V p_\mss^2 + \bar m_V^2 \right]^2} \bar\chi(-p_1) \slashed p_\mss \chi(p_2) \bar\chi(-p_3) \slashed p_\mss \chi(p_4) \\
        		&\quad + \frac{1}{2} \int_{p_2,p_3,p_4} \frac{ \bar h_\phi^2 }{Z_{\phi} p_\mss^2 + \bar m_\phi^2 } \bar\chi(-p_1) \chi(p_2) \bar\chi(-p_3) \chi(p_4) - \frac{1}{2} \int_{p_2,p_3,p_4} \frac{\bar h_V^2}{Z_V p_\mss^2 + \bar m_V^2} \bar\chi(-p_1) \sigma_\mu \chi(p_2) \bar\chi(-p_3) \sigma_\mu \chi(p_4) 
                        \,,
                        \label{eq:17}
\end{aligned}
\end{equation}
\end{widetext}
where $p_\mss = p_3 + p_4$. A comparison to the
action~\eqref{eq:GNTh:EffectiveAverageAction} of the purely fermionic
model reveals that the bosonization procedure induces the Gross-Neveu
and Thirring interactions with an effective $\mss$-channel momentum
dependence.  Moreover, the vector boson parametrizes a further
interaction vertex resembling the square of the kinetic term,
$(\bar\chi \slashed p_\mss \chi)^2$.  This type of interaction is
present if $\bar A_V \neq Z_V$ and is not resolved in our ansatz
\eqref{eq:GNTh:EffectiveAverageAction} for the momentum-dependent
Gross-Neveu-Thirring model.  We come back to this observation
below.

\section{Limiting cases}
\label{sec:LimitingCases}

Let us go back to the purely fermionic case described by the action~\eqref{eq:GNTh:EffectiveAverageAction} and analyze the simplifying limits of pointlike interactions and infinite flavor number, respectively.

\subsection{Pointlike limit}

\begin{figure}[tb]
	\includegraphics[scale=0.9]{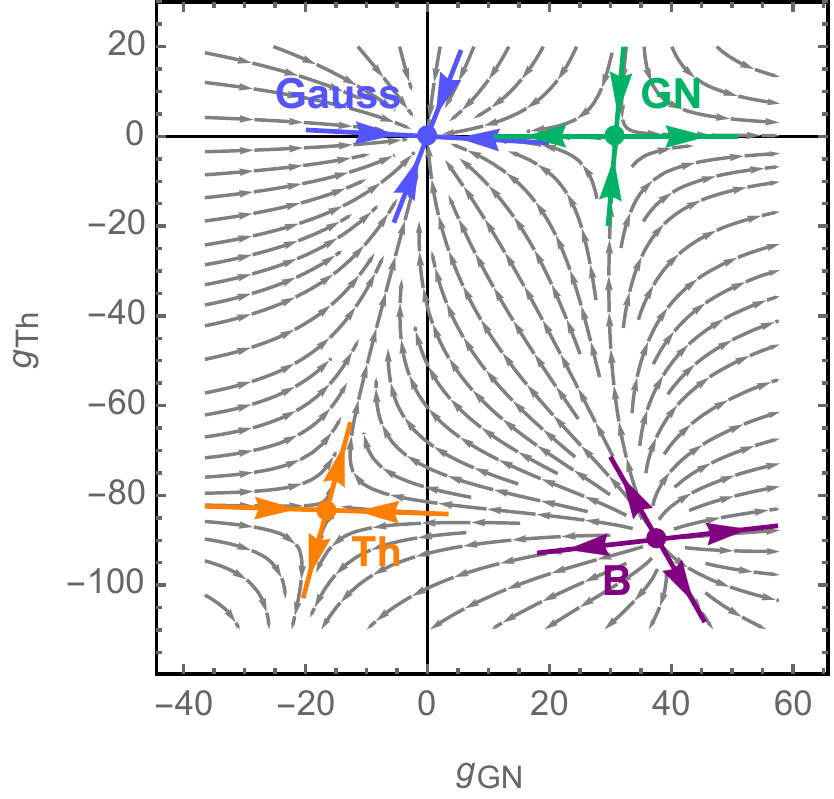} 
\caption{Phase diagram of the Gross-Neveu-Thirring model in the pointlike limit, highlighting the four fixed points, their eigenperturbations, and flow lines towards the \IR\ for $\Nf = 4$. The \fpGN\ fixed point lies on the 
Gross-Neveu axis, whereas the \fpTh\ fixed point approaches the Thirring axes only for $\Nf\to\infty$.}
\label{fig:GNTh:TheorySpacePointlike}
\end{figure}

The RG flow of the Gross-Neveu-Thirring model in the limit of
momentum-independent (pointlike) couplings has been investigated in
detail \cite{Gies:2010st, Janssen_2012, Janssen:2012pq,PhysRevD.88.065008,Jakovac:2014lqa,
  Gehring:2015vja}. This limit forms the basis of derivative
expansions using bosonization techniques to restore parts of the
momentum dependence \cite{PhysRevLett.86.958, Hofling:2002hj,
  Braun:2010tt, Braun:2011pp, Janssen_2012, Janssen:2012pq,
  Janssen:2014gea, Knorr:2016sfs}. In this limit, the self-energy
diagrams~\eqref{eq:DiagramFunctionals:SelfEnergy} do not acquire a
nontrivial momentum dependence, and hence the anomalous dimension
vanishes. Using~\eqref{eq:ThresholdKernel:RelationToThresholdFunction}
the coupling flows reduce to
\begin{subequations}
\label{eq:GNTh:BetaFunction:Pointlike}
\begin{align}
	\partial_t \gGN &= \gGN + \frac{4 v_d}{\tilde N} \ell^{(F)}_1(0; 0) \left[ (-\tilde N + 2) \gGN^2 \right. \notag \\
		&\qquad\qquad\qquad\qquad\qquad \left. + 6 \gGN \gTh + 4 \gTh^2 \vphantom{(-\tilde N + 2) \gGN^2} \right] , 
\label{eq:GNTh:BetaFunction:G:Pointlike} \\
	\partial_t \gTh &= \gTh + \frac{4 v_d}{\tilde N} \ell^{(F)}_1(0; 0) \left[ \frac{\tilde N + 2}{3} \gTh^2 + 2 \gGN \gTh \right], 
\label{eq:GNTh:BetaFunction:V:Pointlike}
\end{align}
\end{subequations}
in accordance, \eg, with Refs.~\onlinecite{Gies:2010st,
  Gehring:2015vja}.  We show the resulting phase diagram in
Fig.~\ref{fig:GNTh:TheorySpacePointlike} for $\Nf=4$ and using
  the Litim regulator \cite{Litim:2001up, Litim:2002cf}.  As
frequently noted in the literature, the pure Gross-Neveu model forms
an invariant subspace in this limit: Setting $\gTh = 0$, we can solve
\Eqref{eq:GNTh:BetaFunction:G:Pointlike} independently of and
consistently with \Eqref{eq:GNTh:BetaFunction:V:Pointlike}. This fact
is used in many studies of the Gross-Neveu model to restrict the
theory space to just the Gross-Neveu vertex, even though, in
principle, also the Thirring vertex has the same symmetries as the
irreducible Gross-Neveu model; in the case of the reducible
Gross-Neveu model, there are even three additional vertices compatible
with the reduced symmetries of the model \cite{Gehring:2015vja}. With
regard to the flow equations~\eqref{eq:GNTh:BetaFunctions} of the
momentum-dependent model, we already observe that this property no
longer holds true in the general momentum-dependent case, because
there are two diagrams of order $\gGN^2$ contributing to the flow of
$\gTh$. We will come back to this point in Sect.~\ref{sec:FiniteNf}.

The phase diagram with its fixed points and flows towards the
long-range physics of the pointlike model has already been discussed
extensively in Ref.~\onlinecite{Gies:2010st}. For our purposes, we
merely recall that the phase diagram exhibits four fixed points as
shown in Fig.~\ref{fig:GNTh:TheorySpacePointlike}. In addition to the
trivial Gaussian fixed point, there are two critical fixed points with
one relevant eigenperturbation, which we will refer to as the
Gross-Neveu (\fpGN) and Thirring (\fpTh) fixed points. While the
\fpGN\ fixed point always lies on the Gross-Neveu axis, the \fpTh\
fixed point approaches the Thirring axis only for $\Nf \to
\infty$. Still, it controls the \IR\ behavior of theories defined on
the negative Thirring axis nonetheless. In addition, there is one more
fixed point having two relevant perturbations and named \fpB\ in the
following.

\subsection{Large-\texorpdfstring{\Nf{}}{Nf} limit}
\label{sec:LimitingCases:LargeNf}

\begin{figure}[tb]
\centering
\includegraphics[scale=0.8]{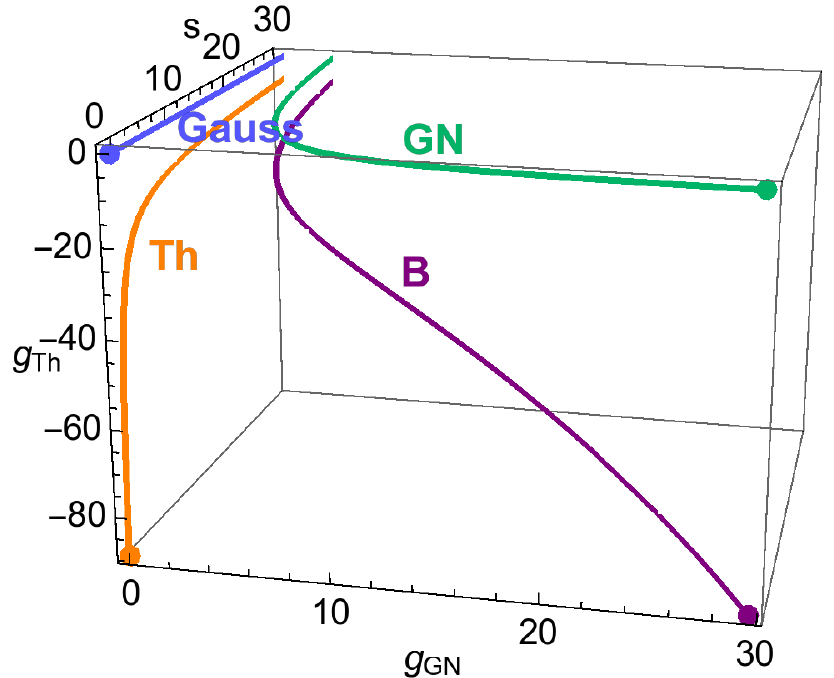}
\caption{Large-$\Nf$ fixed functions of the Gross-Neveu-Thirring model in the $\mss$-channel approximation. The points mark the positions of the fixed points in 
the pointlike limit, which match the fixed-function values at $\mss=0$.}
\label{fig:GNTh:FixedPointSolution:LargeNf}
\end{figure}

As a second instructive limit, we study the fixed-point structure of
the momentum-dependent \RG\ flow of the vertices in the limit $\Nf \to
\infty$, employing the $\mss$-channel approximation. We observe that
the diagram functionals~\eqref{eq:DiagramFunctionals}
and~\eqref{eq:DiagramFunctionals:SelfEnergy} scale with $1/\Nf$ and
hence can be neglected in this limit. In fact, only the $\mss 2$
diagrams in the flow equations for $\gGN$ and $\gTh$ remain finite.
Furthermore, the anomalous dimension~\eqref{eq:GNTh:ADF} vanishes
exactly because $\dgrfSE^2[g; \eta] \equiv 0$ in the $\mss$-channel
approximation. As a result, the flow
equations~\eqref{eq:GNTh:BetaFunction:G}
and~\eqref{eq:GNTh:BetaFunction:V} decouple, and both take the form
\begin{equation}
	\partial_t g(\mss) = g(\mss) + 2 \mss \, g^\prime(\mss) - 2 \kernelFG{a_1}{a_2}(\mss) \, g(\mss)^2
	\label{eq:GNTh:BetaFunctions:LargeNf}
\end{equation}
with $a_1 = 1$, $a_2 = 0$ for $\gGN$ and $a_1 = -1$, $a_2 = 1$ for $\gTh$. Here, we have defined
\begin{equation}
	\kernelFG{a_1}{a_2}(\mss) := \int_q \kernelFG{a_1}{a_2}(p_3+p_4, q; 0),
\end{equation}
being the integrated threshold kernel~\eqref{eq:ThresholdKernel} for
$\mss$-channel processes. Note that this quantity depends only on the
Mandelstam variable $\mss = (p_3+p_4)^2$. The substantial technical
simplification of the large-$\Nf$ limit arises from the fact that the
integral no longer involves the coupling function $g$. Therefore, the
associated fixed-point equation ($\partial_t g_* = 0$) is an ordinary
differential equation. As its solution is a full function rather
  than a ``fixed point'', we refer to such solutions as \textit{fixed
    functions} in the following. In the present case, the general
  solution is
\begin{equation}
\label{eq:GNTh:FixedPointSolution:LargeNf}
	g_*(\mss) = \frac{ \mss^{-1/2} }{g_\infty + \int_\mss^\infty \d x \; x^{-d/2} \kernelFG{a_1}{a_2}(x) } , \quad g_\infty = \text{const.}
\end{equation}
Here $g_\infty$ represents an \textit{a priori} arbitrary constant,
implying that we obtain a one-parameter family of potential fixed
point solutions. The condition that $g_*(\mss)$ should not acquire
singularities for $\mss>0$ is fulfilled for values $g_\infty > 0$
($g_\infty < 0$) for ${\gGN}_*$ (${\gTh}_*$), because
$\kernelFG{1}{0}$ ($\kernelFG{-1}{1}$) is strictly positive
(negative). In fact, even the value $g_\infty = 0$ leads to a
well-defined coupling function for all $\mss \geq 0$, but its
asymptotic large-$\mss$ behavior changes qualitatively: we
find $g_*(\mss) \sim \mss$ if $g_\infty = 0$, whereas $g_*(\mss) \sim
1/\sqrt{\mss}$ if $g_\infty > 0$ ($g_\infty < 0$). For the Litim
regulator \cite{Litim:2001up, Litim:2002cf} the remaining integral in
\Eqref{eq:GNTh:FixedPointSolution:LargeNf} can even be evaluated
analytically, yielding a solution in closed form. Since the resulting
expression is rather extensive, we simply plot the solutions in
Fig.~\ref{fig:GNTh:FixedPointSolution:LargeNf}. The arrangement of
axes is similar to Fig.~\ref{fig:GNTh:TheorySpacePointlike}, with the
projected third axis exhibiting the $\mss$ dependence. The \fpB\ fixed
function is obtained by combining the nontrivial solutions for ${\gGN}_*$
and ${\gTh}_*$.

The integration constant $g_\infty$ can be fixed by including an
additional piece of information from finite-$\Nf$ results: for this,
we first note that the solution
function~\eqref{eq:GNTh:FixedPointSolution:LargeNf} is not analytic:
in a power series expansion around $\mss = 0$, we also encounter
half-integer powers of $\mss$.  This peculiarity is caused by the
kernel $\kernelF{a_1}{a_2}(\mss)$ regardless of the choice of
regulator as is discussed in
Appendix~\ref{app:NonanalyticCouplingFunctions}. It turns out that the
lowest-order nonanalyticity $\sim \sqrt{\mss}$ is absent for
finite-$\Nf$ solutions. This motivates to impose a ``maximum
regularity condition'' also on the large-$\Nf$ solution: we require
the constant $g_\infty$ to be fixed such that as many derivatives of
$g_*$ in $\mss = 0$ exist as possible. This regularity condition
implies $g_\infty = \frac{1}{16}$ for ${\gGN}_*$ and $g_\infty =
-\frac{1}{32}$ for ${\gTh}_*$, which corresponds also to the values
used for the plots in Fig.~\ref{fig:GNTh:FixedPointSolution:LargeNf}.

In order to characterize the fixed point, we proceed by
calculating their spectrum of critical exponents. For this, we perturb
around the fixed function with an ansatz $g(\mss) = g_*(\mss) +
\e^{-\theta t} \varepsilon(\mss)$. Linearizing the flow
equation~\eqref{eq:GNTh:BetaFunctions:LargeNf} in $\varepsilon$ yields
the eigenvalue equation
\begin{align}
	-\theta \varepsilon(\mss) = \varepsilon(\mss) + 2 \mss \, \varepsilon^\prime(\mss) - 4 \kernelFG{a_1}{a_2}(\mss) \, g_*(\mss) \varepsilon(\mss) 
\,.
\end{align}
The resulting eigenperturbations are
\begin{equation}
\label{eq:GNTh:Eigenperturbations:LargeNf}
	\varepsilon_0(\mss) = C \, \mss^{\frac{-1-\theta}{2}}
	\quad \text{and} \quad
	\varepsilon(\mss) = C \, g_*(\mss)^2 \, \mss^{\frac{1-\theta}{2}} ,
\end{equation}
for the Gaussian and nontrivial fixed functions, respectively, with
$C$ being a normalization constant in both cases. In order for the
perturbations to be well-defined in $\mss = 0$, the spectrum must be
bounded from above by $\Re \theta \leq -1$ for the Gaussian and $\Re
\theta \leq 1$ for the nontrivial fixed points. If we additionally
demand for maximum regularity, the critical exponents become quantized
with spacing $2$, \ie\ $\theta = -1, -3, -5, \ldots$ for perturbations
around Gaussian and $\theta = 1, -1, -3, \ldots$ for perturbations
around nontrivial solutions. In fact, the importance of a proper
choice of the Hilbert space for spanning the fixed functions is also
well studied for the analysis of fixed potentials in field space for
Wilson-Fisher type fixed points \cite{Morris:1994ie,Morris:1998da,Rosten:2010vm} and beyond \cite{Bridle:2016nsu,Eichhorn:2016hdi,Morris:2018mhd,Gies:2015lia,Gies:2016kkk}. We observe a similar
feature here for fixed functions in momentum space in the form of the
maximum regularity criterion.

Since we can perturb in the $\gGN$ and $\gTh$ directions of theory
space independently, the resulting spectrum of any of the four fixed
functions is obtained by merging one spectrum for either direction,
leaving the Gaussian fixed point with zero relevant exponents, the
\fpGN\ and \fpTh\ fixed points with one relevant exponent $\theta_1 =
1$ each, and the \fpB\ fixed point with two relevant exponents
$\theta_1 = \theta_2 = 1$. Incidentally, this is also the spectrum
found in the partially bosonized language \cite{Braun:2010tt,
  Janssen:2012pq} as well as a generalized fermionic truncation scheme
\cite{Jakovac:2014lqa}.

\section{Fixed functions and spectra for finite flavor numbers}
\label{sec:FiniteNf}

With the information from the bosonized version and two limiting cases
of the fermionic formulation, we are now equipped to study the
momentum-dependent system for finite $\Nf$. The situation is
considerably more complex because all four types of diagram
functionals contribute to the flow
equations~\eqref{eq:GNTh:BetaFunctions}, and all involve integrals of
the unknown functions $\gGN$, $\gTh$, and $\eta$. In view of the
complications posed by these nonlinear integro-differential equations,
we continue to rely on the $\mss$-channel approximation, and resort to
numerical methods to compute fixed functions in theory space and their
associated spectra.

\subsection{Solution strategy}
\label{sec:FiniteNf:SolutionStrategy}

\begin{figure*}[tb]
\begin{minipage}[t]{\textwidth}
\begin{minipage}[t]{0.245\textwidth}
\flushleft\textbf{(a)}\ \\
\centering
	\includegraphics[scale=0.8]{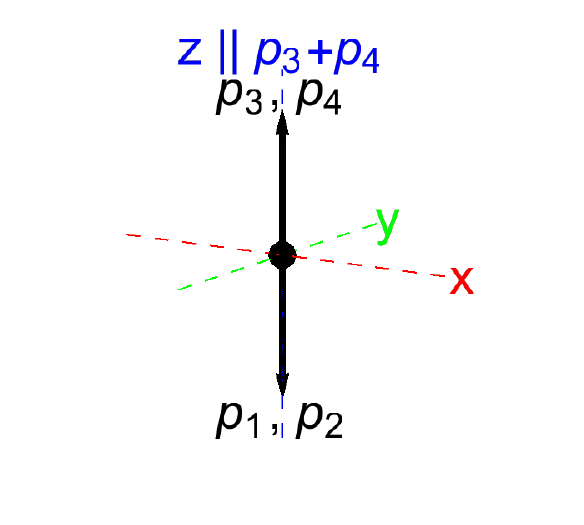}
\end{minipage}	
\begin{minipage}[t]{0.245\textwidth}
\flushleft\textbf{(b)}\ \\
\centering
	\includegraphics[scale=0.8]{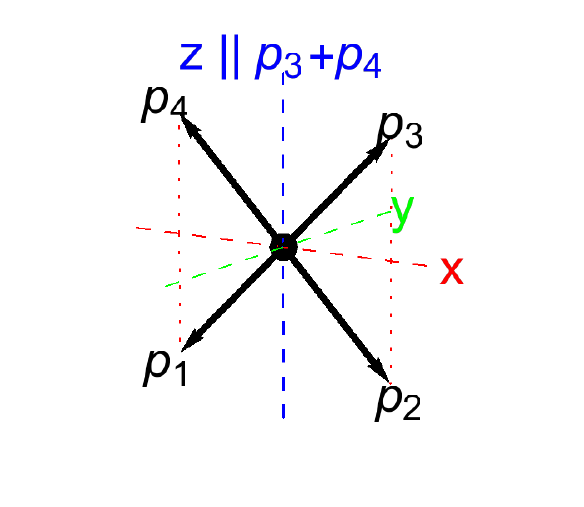}
\end{minipage}
\begin{minipage}[t]{0.245\textwidth}
\flushleft\textbf{(c)}\ \\
\centering
	\includegraphics[scale=0.8]{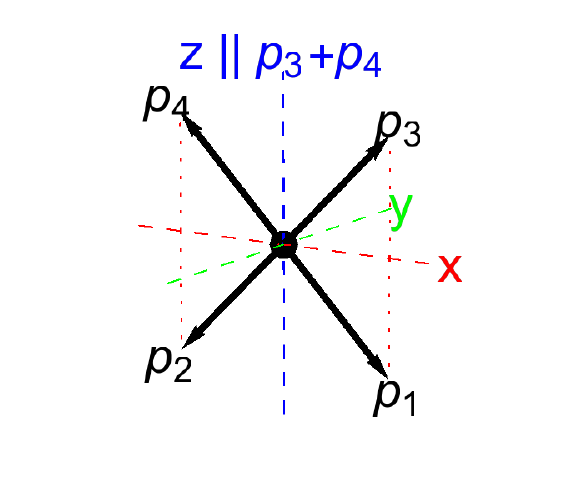}
\end{minipage}
\begin{minipage}[t]{0.245\textwidth}
\flushleft\textbf{(d)}\ \\
\centering
	\includegraphics[scale=0.8]{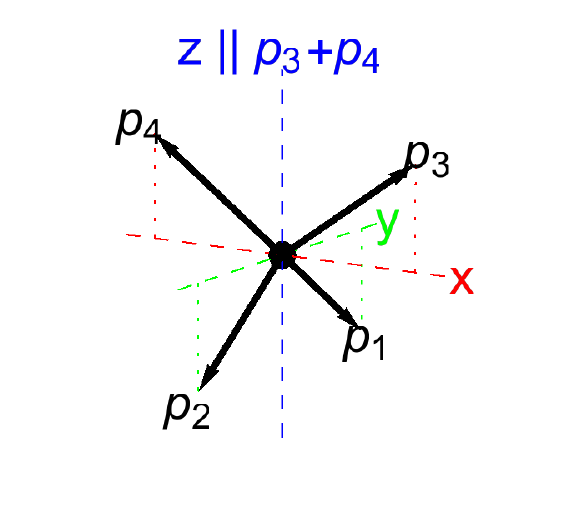}
\end{minipage}
\end{minipage}
\caption{Momentum configurations used in the computation of loop integrals in the $\mss$-channel approximation. All momenta are assumed to have equal magnitude. In 
the parallel configuration (a), the momenta are aligned along the same direction. In the two orthogonal configurations (b) and (c), all momenta lie in one 
plane (the $x$-$z$ plane), and one of the Mandelstam variables $\mst$ or $\msu$ vanishes. We refer to (b), where $\msu = 0$, as the $\mst$-favored, and to (c) 
with $\mst = 0$ as the $\msu$-favored orthogonal configuration. In the symmetric configuration (d), the angle between any two momentum vectors is the same, 
known as the tetrahedral angle $\arccos(-\frac{1}{3}) \approx 109.47\degree$.}
\label{fig:MomentumConfigurations}
\end{figure*}

We first consider the fixed point equations, \ie\ $\partial_t \gGN
= \partial_t \gTh = 0$ in \Eqref{eq:GNTh:BetaFunctions}.  The
integro-differential structure of the equation and the nonanalyticity
of the threshold kernel constitute rather challenging premises for any
numerical computation. Notwithstanding, the pseudo-spectral method
using Chebyshev rational functions \cite{Boyd:ratCheb, Boyd:ChebyFourier,
  Guo:2002crs} has proved to be a versatile and robust tool to obtain
a solution of arbitrary accuracy, at least in principle. As an example
from the \FRG\ context, it has been used successfully to obtain
high-accuracy solutions of effective potentials for various model
systems before \cite{Borchardt:2015rxa, Heilmann:2014iga,
  Borchardt:2016pif, Knorr:2016sfs, Borchardt:2016kco,
  Borchardt:2017fcm, Knorr:2017yze, Knorr:2017mhu}. We will therefore
expand the couplings and the anomalous dimension function as a series
of Chebyshev rational functions $R_n$ ($n = 0, 1, \ldots$). These are
obtained from the Chebyshev polynomials of the first kind $T_n$ by a
compactification of the semi-infinite interval $[0, \infty)$,
\begin{equation}
\label{eq:ChebyshevRationalFunctions}
	R_n(x) := T_n\!\left(\frac{x-L}{x+L}\right) .
\end{equation}
The parameter $L$ is arbitrary, but fixed. It should typically reflect
the order of the length scales of the function to be modeled
\cite{Boyd:ChebyFourier}; for our purposes, $L = 5$ turned out to
yield satisfying convergence. By definition, the Chebyshev rational
functions inherit useful properties such as
analyticity, boundedness, and orthogonality from their polynomial
parents. Most importantly, they form a basis set of $\mathcal
L^2\left([0, \infty) \right)$ with measure $\d x \sqrt{L/x} / (x+L)$,
and the Chebyshev series is guaranteed to converge on the entire
domain if the expanded function is square-integrable
\cite{MasonHandscomb:ChebyshevPolynomials}. This should be contrasted, for
instance, with Taylor series, which converge only within a disc up to the
closest singularity in the complex plane.  Following this approach, we
write the coupling functions $\gGN$, $\gTh$ and the anomalous
dimension function $\eta$ as truncated series of Chebyshev rational
functions,
\begin{equation}
\label{eq:FixedPointSolutions:ChebyshevAnsatz}
	g_*(\mss) = \sum_{n=0}^{n^g_{\max}} g_n R_n(\sqrt{\mss}) \,,
	\quad
	\eta_*(p^2) = \sum_{n=0}^{n^\eta_{\max}} \eta_n R_n(p^2) \,.
\end{equation}
Note that we parametrize the argument for the coupling functions as
$\sqrt{\mss}$ rather than $\mss$. This is motivated by the asymptotic
behavior as $\mss\to\infty$: If the fixed point coupling functions are
bounded, the quantum terms in \Eqref{eq:GNTh:BetaFunctions} are
suppressed by at least $\frac{1}{\sqrt{\mss}}$ compared to the scaling
terms due to the asymptotics of the threshold
kernel~\eqref{eq:ThresholdKernel}. Similarly, the anomalous dimension
function~\eqref{eq:GNTh:ADF} vanishes in the limit $p^2 \to
\infty$. Consequently, we find that $g_*(\mss) \sim
\frac{1}{\sqrt{\mss}}$ as $\mss \to \infty$ for both $\gGN$ and
$\gTh$. The ansatz~\eqref{eq:FixedPointSolutions:ChebyshevAnsatz} is
therefore better adapted to the asymptotic behavior than a plain
$\mss$ parametrization. This turns out to be crucial for achieving
convergence of the numerical solutions.
In particular, a simple parametrization of the
coupling functions similar to \Eqref{eq:17} does not provide a reliable
estimate of their momentum dependence.
Moreover, we already observed
in the previous section that the threshold kernel generates
half-integer powers of $\mss$, so that an expansion of $g_*$ of the
above form appears natural. By contrast, the asymptotic behavior of the
anomalous dimension $\eta_*$, following from \Eqref{eq:GNTh:ADF} together with 
the asymptotics of the couplings, is given by $\eta_*(p^2) \sim \frac{1}{p^2}$
for $p^2 \to \infty$. This justifies the parametrization
of \Eqref{eq:FixedPointSolutions:ChebyshevAnsatz}.

The remaining task is to determine the expansion coefficients
$\gGNCoeff{n}$, $\gThCoeff{n}$, and $\eta_n$ using the pseudo-spectral
method \cite{Boyd:ChebyFourier}. Plugging the series
expansions~\eqref{eq:FixedPointSolutions:ChebyshevAnsatz} into the
relations~\eqref{eq:GNTh:BetaFunctions} and~\eqref{eq:GNTh:ADF} and
demanding the equations to hold on a discrete set of collocations
points $\lbrace \mss_m \rbrace_{m=0}^{n^g_{\max}}$ and $\lbrace p^2_m
\rbrace_{m=0}^{n^\eta_{\max}}$ leaves us with a nonlinear system of
algebraic equations for precisely these coefficients.  To solve this
system, we use a Newton-Raphson method
\cite{AtkinsonHan:TheoreticalNumericalAnalysis,
  Deuflhard:NewtonMethodsForNonlinearProblems} to iteratively reduce
the residuals starting from some initial guess for the coefficients.
For the collocation points, we choose the roots of the next-order
expansion function $R_{n_{\max}+1}$. This ensures the pseudo-spectral
method to coincide with the spectral method if the inner products in
the latter are evaluated numerically using an optimal Gaussian
quadrature rule \cite{Boyd:ChebyFourier}.

For the concrete computation, there is one subtlety which remains to
be addressed: Even after a restriction to a pure $\mss$-channel
dependence, the flow equations~\eqref{eq:GNTh:BetaFunctions} for the
coupling functions do not uniquely determine the remaining $\mss$
dependence of the couplings. For a given value of $\mss$, the
directions of the external momenta in three dimensional spacetime are
not uniquely fixed. In order to compute the corresponding integrals,
we have to make a choice for these directions, and the result for the
$\mss$-dependent coupling does generically depend on this choice. Our
strategy now is to perform the computations for a set of distinct
geometric configurations of momenta, monitoring the variation of the
resulting solution functions upon changing the configuration.  In
fact, this serves as a self-consistency check for the $\mss$-channel
approximation: if the different configurations show large variations,
the contributions from the $\mst$ and $\msu$ channels must be sizable.

The set of momentum geometries consists of the four different
momentum configurations shown in
Fig.~\ref{fig:MomentumConfigurations}, which are expected to display
possible variations rather prominently. In each of them, the magnitude
of all four momenta is the same, $p_i^2 = p^2 = \frac{\mss}{2 + 2
  \cos\sphericalangle(p_3, p_4)}$. In the parallel configuration in
panel (a), we have $\mst = \msu = 0$, leading to the ``purest''
$\mss$-dependence. The orthogonal configurations in panels (b) and (c)
accentuate one of the $\mst$- or $\msu$-channels against the other: In
the $\mst$-favored configuration (b), we have $\mst = \mss$ and $\msu
= 0$, whereas in the $\msu$-favored one (c), $\mst = 0$ and $\msu =
\mss$. In the symmetric configuration in panel (d), the angle between
any two momentum vectors is the same, \ie\ $\cos\sphericalangle(p_i,
p_j) = -\frac{1}{3}$ ($i \neq j$), hence $\mss = \mst = \msu$.

In order to compute the spectra of fixed point solutions, we proceed
similarly to the large-$\Nf$ case discussed above by perturbing around
a fixed function solution with $\gGN(\mss) = \gGNCoeff{*}(\mss) +
\e^{-\theta t} \subGN{\varepsilon}(\mss)$ and $\gTh(\mss) =
\gThCoeff{*}(\mss) + \e^{-\theta t}
\subTh{\varepsilon}(\mss)$. Contrary to the large-$\Nf$ case, however,
we cannot consider perturbations in the $\gGN$ and $\gTh$ directions
independently because the equations do not decouple. Expanding to
first order in $\varepsilon$, we arrive at the linearized flow
equations
\begin{widetext}
\begin{subequations}
\label{eq:GNTh:LinFlowEquations}
\begin{align}
\label{eq:GNTh:LinFlowEquation:G}
&\begin{aligned}
	-\theta \subGN{\varepsilon} &= \left[ 1 + 2 \eta^{(4)} + 2 \mss \, \partial_\mss \right] \subGN{\varepsilon} - \df{\mss 2}{\tilde 
N}{0}_\leftrightarrow[\gGNCoeff{*}, \subGN{\varepsilon}] + \df{\mss 1}{2}{0}_\leftrightarrow[\gGNCoeff{*}, \subGN{\varepsilon}] + \df{\mss 
1}{3}{0}_\leftrightarrow[\gGNCoeff{*}, \subTh{\varepsilon}] + \df{\mss 1}{3}{0}_\leftrightarrow[\subGN{\varepsilon}, \gThCoeff{*}] \\
		&\qquad + \df{\mst}{1}{0}_\leftrightarrow[\gGNCoeff{*}, \subTh{\varepsilon}] + \df{\mst}{1}{0}_\leftrightarrow[\subGN{\varepsilon}, 
\gThCoeff{*}] + \df{\mst}{2}{0}_\leftrightarrow[\gThCoeff{*}, \subTh{\varepsilon}] - \df{\msu}{1}{0}_\leftrightarrow[\gGNCoeff{*}, \subTh{\varepsilon}] - 
\df{\msu}{1}{0}_\leftrightarrow[\subGN{\varepsilon}, \gThCoeff{*}] + \df{\msu}{2}{0}_\leftrightarrow[\gThCoeff{*}, \subTh{\varepsilon}] \,, \\
\end{aligned} \\
\label{eq:GNTh:LinFlowEquation:V}
&\begin{aligned}
	-\theta \subTh{\varepsilon} &= \left[ 1 + 2 \eta^{(4)} + 2 \mss \, \partial_\mss \right] \subTh{\varepsilon} + \df{\mss 2}{\tilde N}{-\tilde 
N}_\leftrightarrow[\gThCoeff{*}, \subTh{\varepsilon}] - \df{\mss 1}{1}{-1}_\leftrightarrow[\gGNCoeff{*}, \subTh{\varepsilon}] - \df{\mss 
1}{1}{-1}_\leftrightarrow[\subGN{\varepsilon}, \gThCoeff{*}] + \df{\mss 1}{2}{-2}_\leftrightarrow[\gThCoeff{*}, \subTh{\varepsilon}] \\
		&\qquad + \df{\mst}{0}{1/2}_\leftrightarrow[\gGNCoeff{*}, \subGN{\varepsilon}] + \df{\mst}{1}{-1/2}_\leftrightarrow[\gGNCoeff{*}, 
\subTh{\varepsilon}] + \df{\mst}{1}{-1/2}_\leftrightarrow[\subGN{\varepsilon}, \gThCoeff{*}] + \df{\mst}{2}{1/2}_\leftrightarrow[\gThCoeff{*}, 
\subTh{\varepsilon}] \\
		&\qquad - \df{\msu}{0}{1/2}_\leftrightarrow[\gGNCoeff{*}, \subGN{\varepsilon}] + \df{\msu}{1}{-1/2}_\leftrightarrow[\gGNCoeff{*}, 
\subTh{\varepsilon}] + \df{\msu}{1}{-1/2}_\leftrightarrow[\subGN{\varepsilon}, \gThCoeff{*}] - \df{\msu}{2}{1/2}_\leftrightarrow[\gThCoeff{*}, 
\subTh{\varepsilon}] \,.
\end{aligned}
\end{align}
\end{subequations}
\end{widetext}
In order to solve these equations and extract the eigenvalues $\theta$, we again use
a pseudo-spectral approach and expand the perturbations in terms of
Chebyshev rational functions. It is instructive to first examine the
asymptotic behavior of the eigenperturbations for $\mss \to
\infty$. The diagram functionals are again all suppressed by at least
the decay of the threshold kernel in this limit, so that only the scaling terms
remain. Consequently, we find $\varepsilon(\mss) \sim \mss^a$, where
the asymptotic power $a$ and the critical exponent $\theta$ are
related by the asymptotic scaling relation
\begin{equation}
\label{eq:GNTh:Eigenpert:ScalingRelation}
	\Re \theta + 2 a = -1 \quad (= 2 - d) \,.
\end{equation}
This implies that $a$ and $\theta$ balance each other: The more
irrelevant the perturbation $\varepsilon$ (smaller $\theta$), the
faster it grows as $\mss \to \infty$ (larger $a$). Hence, in order to
probe the irrelevant part of the spectra beyond $\Re \theta = -1$, we
have to allow for asymptotically growing eigenperturbations. Since the
Chebyshev rational functions are bounded, we multiply by a polynomial
in $\mss$, motivating the ansatz
\begin{equation}
\label{eq:GNTh:Eigenpert:Ansatz}
	\varepsilon(\mss) = (\mss + L)^{a_{\max}} \sum_{n=0}^{n_{\max}^\varepsilon} \varepsilon_n R_n(\mss) \,.
\end{equation}
For the choice of the value of $a_{\max}$, we followed two different
approaches. For the first one, we set $a_{\max}$ to a constant
numerical value, thus keeping the maximum growth of the perturbations
fixed. The advantage of this method is that it reduces
Eqs.~\eqref{eq:GNTh:LinFlowEquations} to an algebraic eigenvalue
problem once the equations are evaluated on a collocation
grid.
As a disadvantage, the required expansion order $n^\varepsilon_{\max}$ grows as we want to reach deeper into the irrelevant part of the spectrum.
Since the possible large-$\mss$ scaling $\varepsilon(\mss) \sim \mss^a$ of the ansatz~\eqref{eq:GNTh:Eigenpert:Ansatz} ranges from $a = a_{\max} - n^\varepsilon_{\max}$ to $a = a_{\max}$,
it can only capture perturbations with exponents $\theta$ such that $\Re\theta \in [ -2 a_{\max}-1, 2 n^\varepsilon_{\max} - 2 a_{\max} - 1]$ according to Eq.~\eqref{eq:GNTh:Eigenpert:ScalingRelation}.
In order to retain the relevant part of the spectrum, we therefore have to increase $n^\varepsilon_{\max}$ when increasing $a_{\max}$.
As the strongly irrelevant exponents are not of particular interest anyway, this disadvantage is typically not an issue.

Our second approach uses a
scheme with a running asymptotics, where $a_{\max} =
\frac{-1-\theta}{2}$ is adapted to the eigenvalue $\theta$ according
to the scaling
relation~\eqref{eq:GNTh:Eigenpert:ScalingRelation}. This method is
computationally much more demanding because the eigenvalue
equations~\eqref{eq:GNTh:LinFlowEquations} become nonlinear. In
practice, it turns out that both methods yield identical exponents
within the validity limits of our approximations obtained from varying
the momentum configuration and/or the regularization scheme.

For the actual computations of fixed functions, eigenperturbations and
critical exponents, we have implemented the procedures described above
in a \CPP\ program. The multidimensional numerical integrations were
carried out using the Cuhre algorithm of the {\small CUBA}
library
\cite{Hahn:2005cal, Hahn:2006tcl, Hahn:2014cc0, Hahn:CubaOnline},
which is a deterministic, high-precision integration scheme featuring
globally adaptive subdivisions. As for the regularization, we mostly
used an exponential regulator with shape function 
\begin{equation}
	r(p^2) = \frac{1}{\sqrt{1 - \e^{-p^2}}} - 1 \,.
\end{equation}
Algebraic eigenvalue equations were solved using the \emph{Eigen3} library \cite{eigen}.

\subsection{Fixed point structure}

\begin{figure*}[tb]
\begin{minipage}[t]{\textwidth}
\begin{minipage}[t]{0.69\textwidth}
\flushleft\textbf{(a)}\ \\[-1.0cm]
\centering
	\includegraphics[scale=0.7]{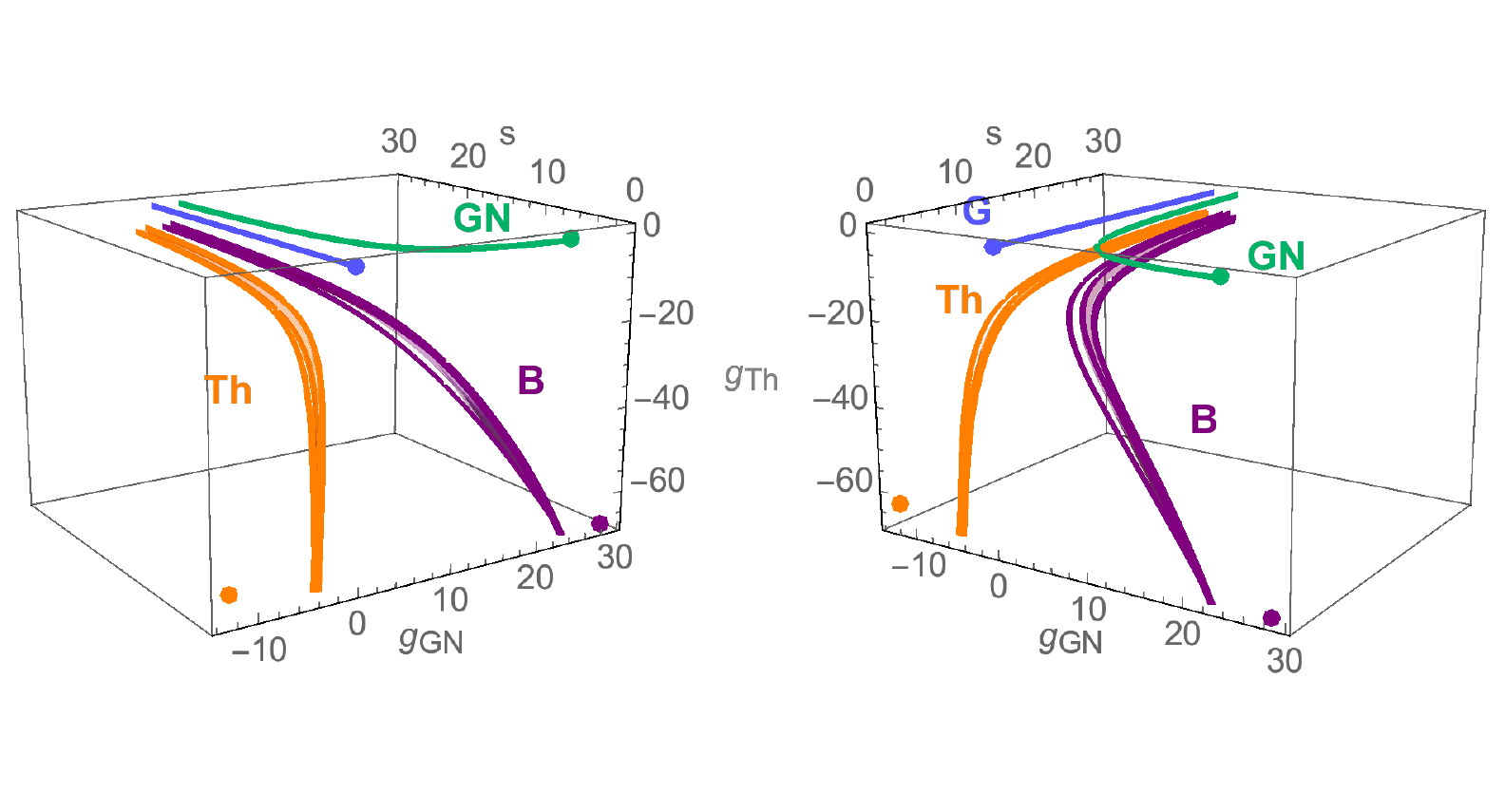}
\end{minipage}	
\begin{minipage}[t]{0.3\textwidth}
\flushleft\textbf{(b)}\ \\[0.75cm]
\centering
	\includegraphics[scale=0.75]{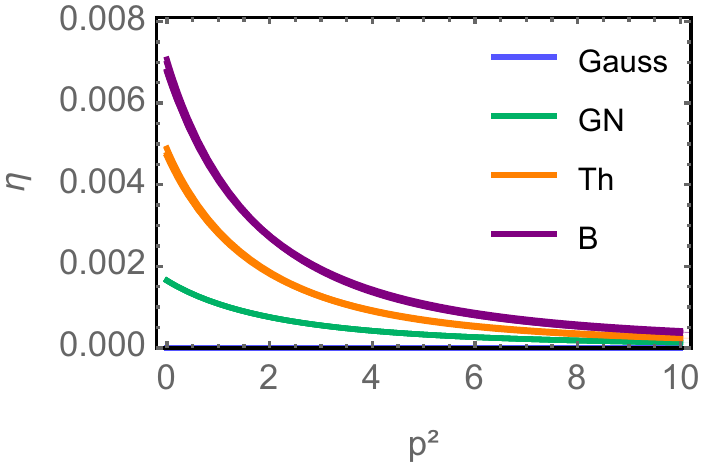}
\end{minipage}
\end{minipage}
\begin{minipage}[t]{\textwidth}
\begin{minipage}[t]{0.69\textwidth}
\flushleft\textbf{(c)}\ \\[-1.0cm]
\centering
	\includegraphics[scale=0.7]{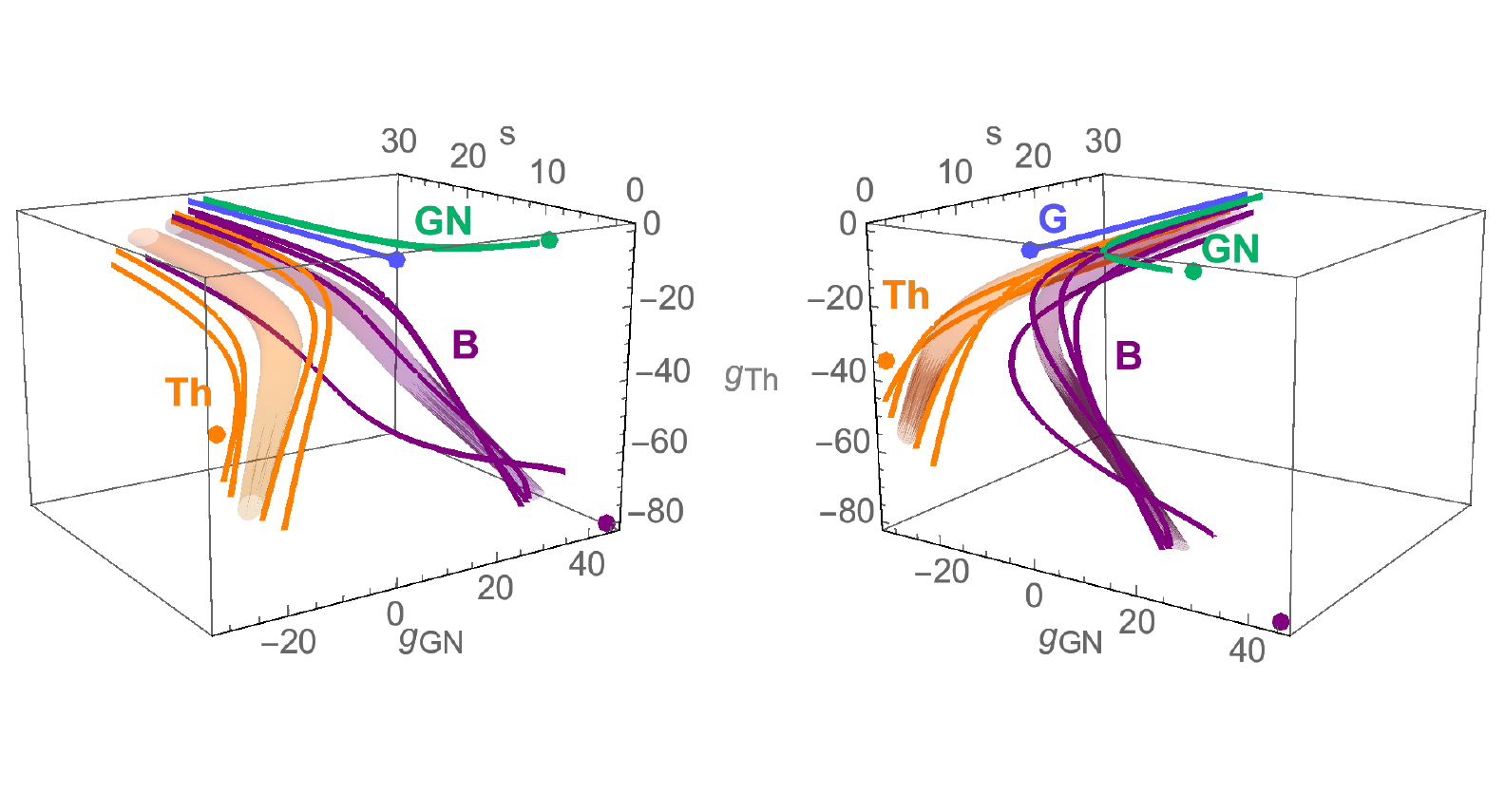}
\end{minipage}	
\begin{minipage}[t]{0.3\textwidth}
\flushleft\textbf{(d)}\ \\[0.75cm]
\centering
	\includegraphics[scale=0.75]{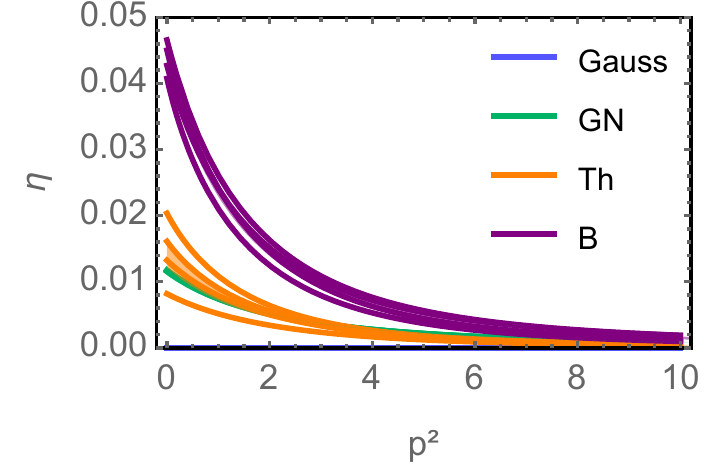}
\end{minipage}
\end{minipage}
\\[-0.5cm]
\caption{Fixed functions for (a-b) $\Nf = 12$ and (c-d) $\Nf = 2$. The
  left panels show the coupling functions in the $\gGN$-$\gTh$ space
  as a function of the Mandelstam variable $\mss$, the right panels
  display the corresponding anomalous dimension functions. The results
  are colored/labeled by their vicinity to the fixed-point
  classification in the pointlike limit $\mss\to 0$. The fixed-point
  values obtained from the pointlike truncation are shown as dots in
  the $\mss=0$ plane. The solid lines for each of these universality
  classes correspond to the four different momentum configurations of
  Fig.~\ref{fig:MomentumConfigurations}. The shaded areas mark their
  mean plus one standard deviation.}
\label{fig:GNTh:FixedPointSolutions}
\end{figure*}

The overall fixed point structure resembles the pointlike and
large-$\Nf$ limits, but there are a few modifications that deserve
attention.  We show a selection of fixed functions for $\Nf=12$ (upper
row) and $\Nf=2$ (lower row) in
Fig.~\ref{fig:GNTh:FixedPointSolutions}.  The actual solutions for the
four momentum configurations of Fig.~\ref{fig:MomentumConfigurations}
are all plotted as solid lines.  To visualize the deviation arising
from the variation of momentum configurations, we additionally display
elliptic tubes around the mean of these configurations with principle
axes given by the standard deviations.  As in the limiting cases
discussed above, we can distinguish four different solutions, the
endpoint of which for $\mss\to 0$ approximately coincides with the
fixed-point positions in the pointlike limit. Correspondingly, we
associate the fixed functions with the Gaussian, the (irreducible)
Gross-Neveu, the Thirring, and the ``B'' universality classes. The
coupling and anomalous dimension functions decay for $\mss \to \infty$
or $p^2 \to \infty$, respectively.

\begin{figure*}[tb]
\begin{minipage}{\textwidth}
\centering
\begin{minipage}[t]{0.325\textwidth}
\flushleft\textbf{(a)}\ \\
\centering
	\includegraphics[scale=0.95]{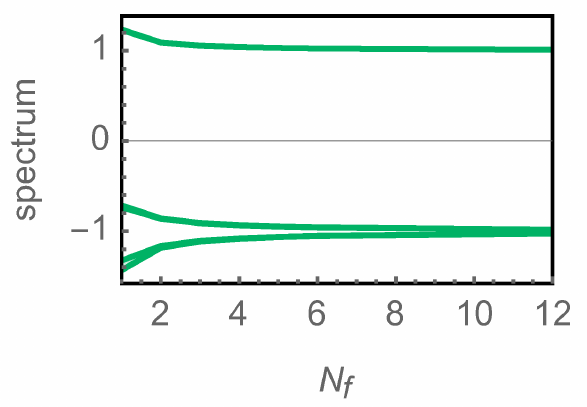}
\end{minipage}	
\begin{minipage}[t]{0.325\textwidth}
\flushleft\textbf{(b)}\ \\
\centering
	\includegraphics[scale=0.95]{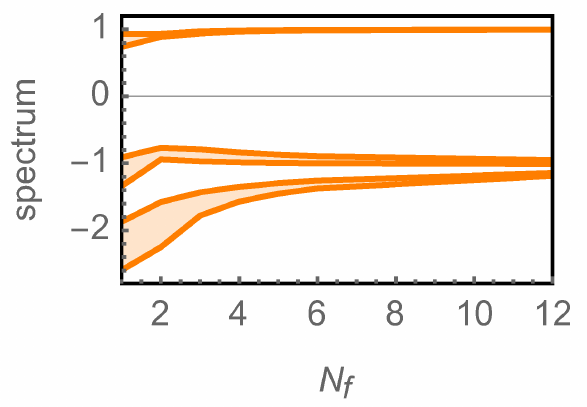}
\end{minipage}
\begin{minipage}[t]{0.325\textwidth}
\flushleft\textbf{(c)}\ \\
\centering
	\includegraphics[scale=0.95]{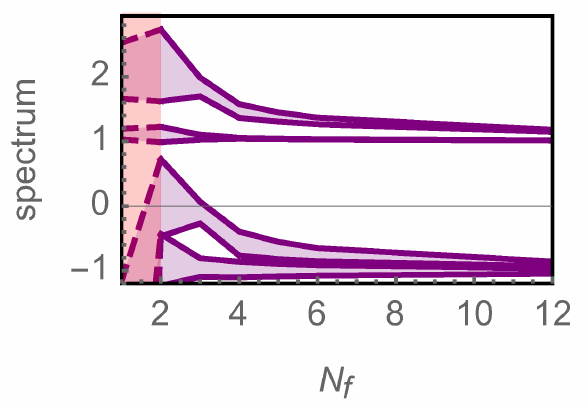}
\end{minipage}
\end{minipage}
\caption{Leading eigenvalues of the (a) \fpGN, (b) \fpTh, and (c) \fpB\ fixed points. The shaded area corresponds to the range spanned by the minimum and maximum value obtained for different momentum configurations. The additional red shading at $\Nf = 1 \ldots 2$ in (c) is to indicate that these results are disputable due to diminished or even missing convergence of fixed point solutions.}
\label{fig:GNTh:CriticalExponents:Leading}
\end{figure*}

For relatively large flavor numbers, \eg\ $\Nf = 12$ in Fig.~\ref{fig:GNTh:FixedPointSolutions}a-b, the variation with momentum configuration is very small for all fixed points and the solutions are close to the infinite-$\Nf$ solution (\cf\ Fig.~\ref{fig:GNTh:FixedPointSolution:LargeNf}).
The $\mss\to0$ limit of the fixed function is also quantitatively close to the fixed-point value in the pointlike limit for the Gross-Neveu universality class, whereas the Thirring and the ``B'' fixed points show some larger deviations.
The agreement with the large-$\Nf$ limit is a manifestation of the fact that the $\mss$-channel approximation is closed in this limit.
For smaller values of $\Nf$, the \fpGN\ fixed function remains remarkably stable when varying the momentum configuration, whereas the deviations increase significantly for the \fpTh\ and \fpB\ fixed points.
In fact, we cannot even make a firm statement about the existence of the \fpB\ fixed point in the current approximation,
because convergence of the Newton-Raphson iteration slowed down significantly here when $\Nf < 2$. In the symmetric configuration, we have even been unable to find a solution.

A plot of the leading eigenvalues of the perturbations around the three nontrivial fixed functions is presented in Fig.~\ref{fig:GNTh:CriticalExponents:Leading}.
For each eigenvalue, we show its minimum and maximum value for different momentum configurations.
These values essentially coincide for the \fpGN\ fixed point, and they vary only moderately and for small $\Nf$ at the \fpTh\ fixed point.
Both of them have one relevant exponent larger than zero as in the pointlike and large-$\Nf$ limits. The \fpB\ fixed point shows stronger variations and apparently also a qualitative change of behavior: a third exponent appears to become relevant in the $\msu$-favored orthogonal configuration for $\Nf = 2, 3$, casting doubt on the reliability of our results in this parameter region.
In any case, the large-$\Nf$ limit is rediscovered consistently, as the eigenvalues approach the values found in Section~\ref{sec:LimitingCases:LargeNf}.
We will now analyze all three nontrivial fixed points in more detail.

\subsection{Gross-Neveu fixed point}

\begin{figure}[tb]
\centering
\includegraphics[scale=0.95]{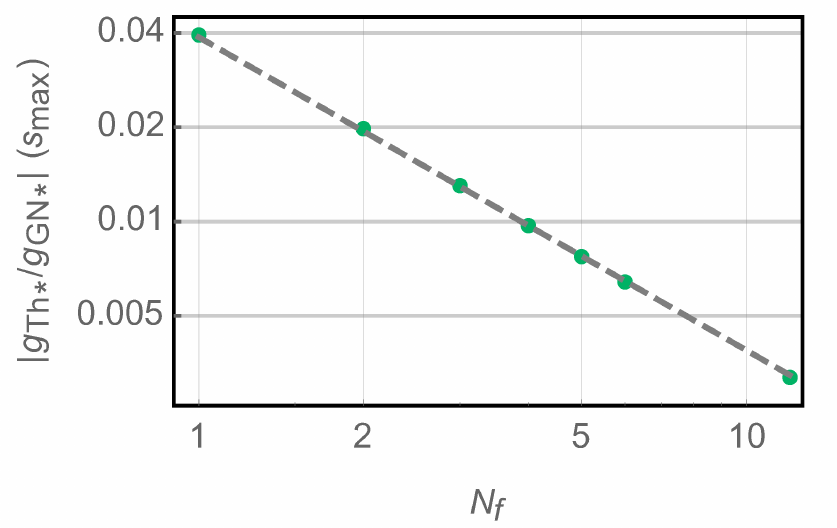}
\caption{Ratio of Thirring and Gross-Neveu couplings at the maximum of the Thirring coupling, measuring the violation of the invariant-subspace property of the Gross-Neveu model, as a function of $\Nf$ (log-log plot).}
\label{fig:GNTh:GNInvariantSubspaceViolation}
\end{figure}

In the pointlike approximation, we found that the Gross-Neveu vertex forms an invariant subspace of theory space: the coupling $\gGN$ does not induce the flow of any of the other symmetry-compatible vertices \cite{Gehring:2015vja}. In particular, no flow of $\gTh$ in Eq.~\eqref{eq:GNTh:BetaFunction:V:Pointlike} is induced on the Gross-Neveu axis.
The flow equations of the fully momentum-dependent model~\eqref{eq:GNTh:BetaFunctions} tell us that this does no longer hold in general. There are $\mst$- and $\msu$-channel processes of order $\gGN^2$ entering the flow of $\gTh$, which simply happen to cancel in the pointlike limit.
Consequently, the Gross-Neveu coupling does not parametrize an invariant subspace any longer, because we cannot find a consistent general fixed function $\gGNCoeff{*}$ after setting $\gTh \equiv 0$ in~Eqs.~\eqref{eq:GNTh:BetaFunction:Pointlike}.
This should serve as a word of caution when restricting the theory space to invariant subspaces in the pointlike limit.
Nevertheless, the violation of this invariance is reassuringly mild in the Gross-Neveu universality class, at least in the $\mss$-channel approximation. For instance in Fig.~\ref{fig:GNTh:FixedPointSolutions}, there is no visible difference.
To illustrate how far the fixed functions actually leave the Gross-Neveu subspace, we compute the maximum value of ${\gTh}_*$ and relate it to ${\gGN}_*$ at the same $\mss$ value.
This ratio, $\left\lvert \frac{{\gTh}_*(\mss_{\max})}{{\gGN}_*(\mss_{\max})} \right\rvert$ with $\mss_{\max}$ such that $\lvert{\gTh}_*(\mss)\rvert \leq \lvert{\gTh}_*(\mss_{\max})\rvert$ for all $\mss$, is plotted in Fig.~\ref{fig:GNTh:GNInvariantSubspaceViolation}.
It reaches up to $4\%$ for $\Nf = 1$ and decays like $1/\Nf$ with the flavor number.
Hence the restriction to $\gGN$ alone remains a satisfactory approximation for all integer $\Nf$.

A precise determination of the critical exponents has remained a quantitative challenge for many computational methods, with results having been accumulated over the past years by numerous approaches, \eg\ various forms of perturbation theory \cite{Luperini:1991sv, Rosenstein:1993zf, Gracey:1993kb, Gracey:1993kc, Gracey:2016mio, Mihaila:2017ble,Zerf:2017zqi,Ihrig:2018hho}, bosonized \FRG\ \cite{PhysRevLett.86.958, Hofling:2002hj, Braun:2010tt, Scherer:2012nn,Vacca:2015nta, Knorr:2016sfs}, conformal bootstrap \cite{Vasiliev:1992wr, Vasiliev:1992wr, Iliesiu:2017nrv}, and Monte Carlo simulations \cite{Karkkainen:1992sk, Karkkainen:1993ef, Chandrasekharan:2013aya, Wang:2014cbw, Li:2014aoa, Hands:2016foa, Schmidt:2016rtz,Huffman:2017swn}. While consensus has been achieved for large $\Nf$ as well as for the case of emergent supersymmetry $\Nf\to1/4$ \cite{Gies:2017tod,Iliesiu:2017nrv,Ihrig:2018hho}, results for the interesting cases of $\Nf=2$ (graphene-like) or $\Nf=1$ (spinless fermions on a honeycomb lattice) have varied rather substantially.

  In fact, our approach, which concentrates on resolving the momentum structure of fermionic four-point vertices, should not be expected to be adequate to produce high-precision estimates for critical exponents. The latter certainly require an accurate treatment of order parameter fluctuations. From large-$\Nf$ computations, it is clear that local order parameter self-interactions contribute to the dynamics near criticality. In the fermionic description, these would correspond to 8-fermion operators which are not included in the present study. For instance, large-$\Nf$ expansions predict that the relevant eigenvalue $\theta_1$ approaches the large-$\Nf$ limit from below, $\theta_1 \simeq 1 - \frac{0.270}{\Nf}$, for $\Nf\to \infty$, whereas our results exhibit an approach to $\theta_1|_{\Nf\to\infty}=1$ from above. Incidentally, the large-$\Nf$ result $\theta_1=1$ can be proved to hold for any critical fixed point in the pointlike limit \cite{Gies:2003dp,Gehring:2015vja}. We observe that this correct limit is also obtained with momentum dependent vertices.

  In order to assess the relevance of the present momentum-resolving approximation, we take a closer look at the critical exponents in the regime of small flavor number $\Nf$, as displayed in Fig.~\ref{fig:GNTh:CriticalExponents:Leading}a. For instance in the case $\Nf=1$, we find $\theta_1\simeq 1.23(3)$, $\eta_\psi\simeq0.028(7)$, where the uncertainties reflect the variation of the result upon choosing different regulators. This can be compared to most recent four-loop $4-\epsilon$ expansion (incl. Borel resummation) \cite{Ihrig:2018hho}, yielding $\theta_1\simeq 1.114(33)$, $\eta_\psi\simeq0.102(12)$, or similarly to FRG results including partial bosonization:  $\theta_1\simeq 1.075(4)$, $\eta_\psi\simeq0.0645$. We observe an agreement on the 10\dots15\% level for $\theta_1$, but a substantially larger deviation for $\eta_\psi$. 

  The picture is similar for the graphene-like case $\Nf=2$, where we obtain  $\theta_1\simeq 1.09(1)$, $\eta_\psi\simeq0.012(2)$, to be compared to the four-loop $4-\epsilon$ expansion (incl. Borel resummation), $\theta_1\simeq 0.993(27)$, $\eta_\psi\simeq0.043(12)$, and the FRG with partial bosonization,  $\theta_1\simeq 0.994(2)$, $\eta_\psi\simeq0.0276$. These results suggest that the class of operators considered in this work do have a satisfactory overlap with those that dominate the leading critical exponent, whereas the anomalous dimension is not well captured. Incidentally, various results from Monte-Carlo simulations and conformal bootstrap show larger deviations (already among each other, as well as with respect to the quoted $\epsilon$ expansion and FRG findings).

\subsection{Thirring fixed point}

As for the \fpTh\ fixed point solution, the variation with momentum configuration is much larger and becomes significant for small flavor numbers. Fig.~\ref{fig:GNTh:FixedPointSolutions} demonstrates that the $\mss$-channel approximation becomes rather crude towards smaller flavor numbers and cannot capture all essential dependencies.
As expected, the outermost \fpTh\ solutions in  Fig.~\ref{fig:GNTh:FixedPointSolutions} correspond to the two orthogonal configurations where the influence of either of the remaining two Mandelstam channels is maximally exposed against the other; the parallel and symmetric configurations lie in between.
The same holds true for the anomalous dimensions, where we observe considerable variation between the configurations for small momenta.
From top to bottom, the curves in both plots belong to the $\mst$-favored orthogonal, symmetric, parallel, and $\msu$-favored orthogonal configurations, respectively.
We conclude that the momentum dependence becomes increasingly important as $\Nf$ becomes smaller, indicating that neither the pointlike limit nor the  $\mss$-channel approximation can be expected to resolve all features.
Nevertheless, the qualitative picture is consistent, and the additional momentum dependence does not modify the topology of the phase diagram. In particular, our results confirm the existence of the Thirring fixed point and thus of a nonperturbatively renormalizable continuum quantum field theory for a wide range of $\Nf$ values. 

This viewpoint is corroborated by our results for the critical exponents displayed in Fig.~\ref{fig:GNTh:CriticalExponents:Leading}b.
Deviations between the momentum configurations become visible only for small $\Nf$ and are stronger in the sub-leading exponents, whose accuracy is naturally lower.
The leading exponent, however, varies only little down to $\Nf = 2$, where the differences affect the second decimal place and are presumably still smaller than those from regulator variations for fixed momentum configuration.
Exponents for various flavor numbers are listed in Tab.~\ref{tab:GNTh:ThCriticalExponents} with error ranges corresponding to the configurational variation bands in Fig.~\ref{fig:GNTh:FixedPointSolutions}.
For the anomalous dimension $\eta_\psi = \eta_*(0)$, the differences between the momentum configurations are larger.
The general trend is again a decreasing value with increasing $\Nf$, eventually converging to $\eta_\psi = 0$ as $\Nf\to\infty$.
The anomalous dimension for $\Nf = 1$, however, cannot reliably be extracted since $\eta_\psi$ flips sign for some momentum configurations in our approximation.

Critical exponents of the Thirring model have also been computed using the \FRG\ with dynamical bosonization for scalar as well as vector boson channels\cite{Janssen:2012pq}.
The values found this way for the relevant exponent $\theta_1$ are generally smaller than our results, \eg, $\theta_1 = 0.4$ for $\Nf = 2$ or $\theta_1 = 0.92$ for $\Nf = 4$,
but the differences reduce as $\Nf$ is increased. Also both \FRG\ results agree on the finding that the large-$\Nf$ limit $\theta_1 = 1$ is approached from below.
In view of our analysis in Section~\ref{sec:PartialBosonization}, a better resolution of the vector channel suggests to include the kinetic term squared, $(\bar\chi \slashed p_\mss \chi)^2$, in future studies. This vertex is naturally entailed in the dynamically bosonized description\cite{Janssen:2012pq} at the same level as the Thirring interaction.

While this analysis of the fixed-point properties provides information about the high-energy behavior of the model and its possible UV completion as an asymptotically safe quantum field theory, many studies in the literature focus on the long-range IR behavior and possible low-energy phases. Here, a variety of earlier studies suggested the existence of a critical flavor number $\Nfcrit$ below which spontaneous breaking of the chiral/flavor symmetry can occur at sufficiently strong coupling\cite{PhysRevD.43.3516, Hong:1993qk, Hands:1994kb, Itoh:1994cr,Kondo:1995np,Hands:1999id, Christofi:2007ye, Gies:2010st,Chandrasekharan:2011mn,Janssen:2012pq}. As discussed in Refs.~\onlinecite{Janssen:2012pq,Gehring:2015vja}, the concrete realization of chiral symmetry in these calculations may have strongly influenced the critical quantities, potentially leading to a different universality class. For instance, the phase transitions observed with lattice realizations using staggered fermions \cite{Kim:1996xza, DelDebbio:1997dv, DelDebbio:1999he, Hands:1999id, Christofi:2007ez, Chandrasekharan:2011mn}, exhibited chiral phase transitions for smaller flavor numbers, \eg, for $\Nf=2$ reporting a critical exponent $\theta_1$ in the range $1.2 \ldots 1.4$. Only recently, lattice simulations have been performed that realize the chiral $\operatorname U(2\Nf)$ symmetry using domain-wall  \cite{Hands:2016foa,Hands:2018vrd} or SLAC \cite{Wellegehausen:2017goy,Lenz:2019qwu} fermions. The results using domain-wall fermions indicate that $0<\Nfcrit<2$ with some preference for $\Nfcrit>1$ by the data  \cite{Hands:2018vrd}.
Latest results of Ref.~\onlinecite{Lenz:2019qwu} using SLAC fermions with an analytic
continuation of the parity-even theory to arbitrary real $\Nf$ provide clear evidence that the critical flavor number is smaller than unity, $\Nfcrit<1$. The latter implies that no
phase transition exists for integer $\Nf$.

\begin{table}[tb]
\caption{Critical exponents with error bounds given by the variation with the external momentum configuration. This provides a lower bound of the error only and does not include deviations between different regularization schemes. In addition, the differences between the distances to the \NJL\ and Thirring subspaces at $\mss = 0$ and $\mss =\infty$ are given.}
\label{tab:GNTh:ThCriticalExponents}
\setlength\extrarowheight{3pt}
\setlength\tabcolsep{4pt}
\begin{tabularx}{\columnwidth}{r|l l|l l}
\hline
\bfseries $\bm{\Nf}$ & \multicolumn{1}{c}{\bfseries $\bm{\theta_1}$} & \multicolumn{1}{c|}{\bfseries $\bm{\eta_\psi}$} & \multicolumn{1}{c}{\bfseries $\bm{\Delta_{\scriptscriptstyle\text{NJL}-\text{Th}}^{\,0}}$} & \multicolumn{1}{c}{\bfseries $\bm{\Delta_{\scriptscriptstyle\text{NJL}-\text{Th}}^{\,\infty}}$} \\
\hline\hline
$1$ & $0.84(10)$ & $0.004(9)$ & $-0.59(11)$ & $-0.67(4)$  \\
$2$ & $0.91(2)$ & $0.014(6)$ & $-0.13(24)$ & $-0.54(16)$   \\
$3$ & $0.95(2)$ & $0.013(2)$ & $\hphantom{-}0.19(12)$ & $-0.36(32)$  \\
$4$ & $0.97(1)$ & $0.012(1)$ & $\hphantom{-}0.34(7)$ & $-0.28(42)$  \\
$5$ & $0.983(4)$ & $0.010(1)$ & $\hphantom{-}0.42(4)$ & $-0.16(42)$  \\
$6$ & $0.989(3)$ & $0.0088(4)$ & $\hphantom{-}0.47(3)$ & $-0.06(39)$   \\
$12$ & $0.9967(3)$ & $0.0047(1)$ & $\hphantom{-}0.59(1)$ & $\hphantom{-}0.27(25)$  \\
\hline
\end{tabularx}
\end{table}

\begin{figure}[tb]
\centering
	\hspace{1cm}\raisebox{6pt}{$\mathsf{\bm{\Nf}}$} \includegraphics[scale=0.85]{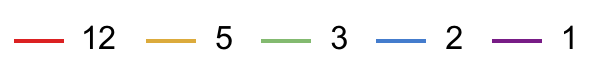} \\
	\includegraphics[scale=0.85]{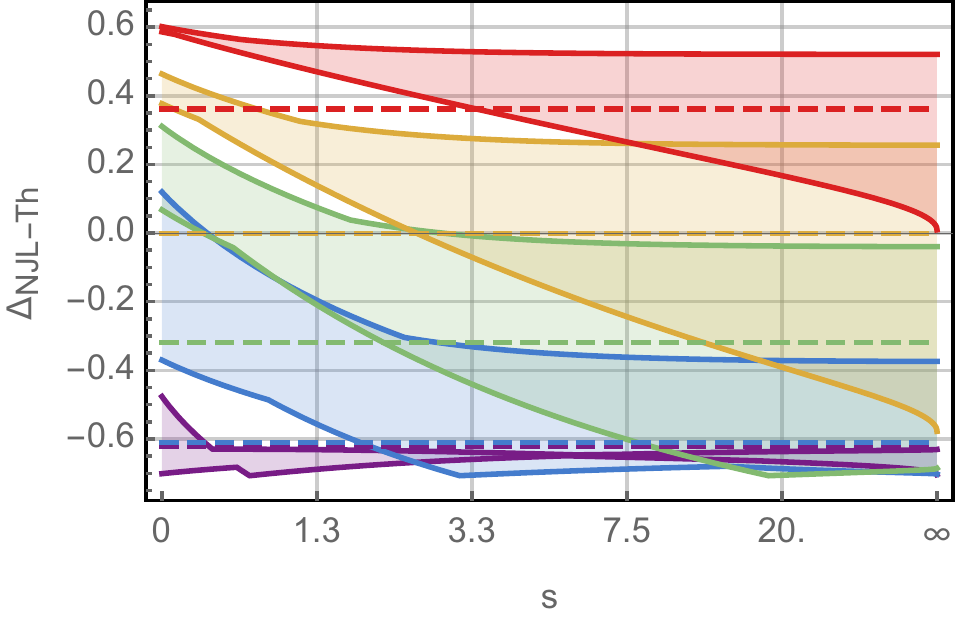}
\caption{Comparison of the distance of the \fpTh\ fixed point solution from the \NJL\ and Thirring subspaces for various flavor numbers. The shaded areas depict the variation among momentum configurations. Note that the kinks in the boundary are caused by the absolute value in the definition of $\subNJL{d}$ and the fact that the configuration with minimum/maximum value for $\Delta_{\scriptscriptstyle\text{NJL}-\text{Th}}$ changes with $\mss$. The dashed lines indicate the values found in the pointlike limit.}
\label{fig:GNTh:FixedPoints:dNJL:NfCompare}
\end{figure}

The present exploratory study should not be expected to give accurate estimates for the critical flavor number. The observation of an increasing relevance of momentum dependence at small $\Nf$ indicates that a larger operator basis is needed to extract the critical quantities reliably. Nevertheless, we can use the range of uncertainties of the present calculation to estimate a range of possible values for $\Nfcrit$. For this, we use
the observation made in Refs.~\onlinecite{Gies:2010st, Janssen:2012pq} that the \fpTh\ fixed point moves from the Thirring subspace towards an \NJL-type subspace as $\Nf$ decreases.
This suggests to interpret the presence or absence of chiral symmetry breaking as a competition between the \NJL\ and Thirring interactions, where the former supports but the latter suppresses the formation of a chiral condensate (as suggested by the large-$\Nf$ limit \cite{PhysRevD.43.3516, Hands:1994kb}).
In the irreducible representation, the \NJL\ vertex is given by the scalar flavor nonsinglet \cite{Janssen:2012pq}
\begin{equation*}
	\gNJL(p_1, p_2, p_3, p_4) \, \bar\chi^i(-p_1) \chi^j(p_4) \bar\chi^j(-p_3) \chi^i(p_2) \,.
\end{equation*}
By a Fierz transformation, we can relate the coupling function $\gNJL(p_i)$ to the Gross-Neveu and Thirring couplings and obtain
\begin{equation}
\begin{aligned}
	\gNJL(p_1, p_2, p_3, p_4) &= -\frac{1}{2} \left[ \gGN(p_1, p_2, p_3, p_4) \right. \\
		&\quad\qquad \left. + \gTh(p_1, p_2, p_3, p_4) \right]
\end{aligned}
\end{equation}
with the above specified convention for the external momenta.
This implies that the \NJL\ subspace is characterized by $\gGN = \gTh$.
In the pointlike limit, the fixed-point coordinates of the Thirring fixed point fulfill this condition for $\Nf = 1.75$ \cite{Janssen:2012pq}.
As a measure for the distance of a point $(\gGN, \gTh)$ in theory space to the \NJL\ subspace, we use
\begin{equation}
	\subNJL{d} := \frac{\left\lvert  \gTh - \gGN \right\rvert}{\sqrt{2}} \,,
\end{equation}
which is a function of the external momenta or, in our case, the Mandelstam variable $\mss$.
In a similar fashion, we can define the distance from the Thirring subspace as $\subTh{d} := \lvert \gGN \rvert$.
We assess the competition between the Thirring and \NJL\ interactions by comparing these distances in Fig.~\ref{fig:GNTh:FixedPoints:dNJL:NfCompare}.
More precisely, we plot the difference $\subNJL{d} - \subTh{d}$ normalized by the location of the fixed point as a function of $\mss$:
\begin{equation}
	\Delta_{\scriptscriptstyle\text{NJL}-\text{Th}} := \frac{\subNJL{d} - \subTh{d}}{\sqrt{\gGN^2 + \gTh^2}} .
\end{equation}
As before, we indicate the variation between the different momentum configurations by a shaded area.
Negative values of $\Delta_{\scriptscriptstyle\text{NJL}-\text{Th}}$ mean that the solution is closer to the \NJL\ subspace, whereas for positive values it is closer to the Thirring subspace.
The dashed lines in the figure show the values of this difference as found in the pointlike limit.
In this case, the value $\Nf = 5.01$ where $\subNJL{d} = \subTh{d}$ lies very close to the critical value $\Nfcrit = 5.1(7)$ reported in Ref.~\onlinecite{Janssen:2012pq} as obtained from a derivative expansion monitoring the chiral order parameter $\langle \bar\psi \psi \rangle$ directly.

The $\mss$-channel momentum dependence reveals that the fixed point solutions tend to be closer to the Thirring subspace for small values of $\mss$.
This suggests $\Nfcrit$ to be smaller than the naive pointlike estimate and thus also of the more elaborate result based on the derivative expansion \cite{Janssen:2012pq}.
Extracting an improved estimate is inhibited by the variations of $\Delta_{\scriptscriptstyle\text{NJL}-\text{Th}}$ with the momentum configuration which increase substantially as $\mss\to\infty$.
As $\Delta_{\scriptscriptstyle\text{NJL}-\text{Th}}$ is a relative measure, this increase can be attributed in part to the fact that $\gGN, \gTh \to 0$ as $\mss \to \infty$.
Explicit estimates and variations for $\Delta_{\scriptscriptstyle\text{NJL}-\text{Th}}$ at $\mss = 0$ and $\mss = \infty$ are also listed in Tab.~\ref{tab:GNTh:ThCriticalExponents}.

Still, we can use  $\subNJL{d} \lesssim \subTh{d}$, or $\Delta_{\scriptscriptstyle\text{NJL}-\text{Th}}\lesssim 0$, as an indicator for chiral symmetry breaking to occur.
Requiring this condition for \emph{some} $\mss$ and \emph{all} momentum configurations would lead to $3 < \Nfcrit < 4$. If we demanded that it holds at $\mss = 0$ for \emph{some} momentum configuration, we would find $2 < \Nfcrit < 3$. Requiring it for \emph{all} $\mss$ and \emph{all} momentum configurations gives $\Nfcrit<2$. The result of this last and most strict criterion is compatible with those of recent lattice simulations \cite{Hands:2018vrd,Lenz:2019qwu}.

In the literature, it has been conjectured that the chiral-symmetry breaking phase transition of the 2+1 dimensional Thirring model and that of QED${}_3$ could be related \cite{Itoh:1994cr,DelDebbio:1999he} or even in the same universality class \cite{Braun:2014wja}. If so, our result based on the most strict criterion is also compatible with recent lattice simulations\cite{Karthik:2015sgq,Karthik:2016ppr} of QED${}_3$.

\subsection{B fixed point}

Contrary to the \fpGN\ and \fpTh\ fixed points, convergence of the fixed point couplings worsens significantly for small flavor numbers at the \fpB\ fixed point.
This manifests itself in two ways: On the one hand, we needed to fine-tune the initial guesses provided to the fixed point solver much more thoroughly to achieve convergence of the Newton-Raphson iteration.
On the other hand, the residuals of the obtained solution functions between the collocation points increase strongly at fixed order of the Chebyshev expansions when $\Nf$ is lowered.
These trends are absent at the \fpGN\ and \fpTh\ fixed points.

At the same time, we notice increasing deviations between the different momentum configurations, similar to the \fpTh\ fixed point.
Furthermore, the \fpB\ fixed point solutions appear to merge into the \fpTh\ solutions for small flavor numbers as $\mss\to\infty$ (see, \eg, Fig.~\ref{fig:GNTh:FixedPointSolutions}c).
In some configurations, the Gross-Neveu coupling $\gGN$ crosses over from $\gGN > 0$ to $\gGN < 0$ as $\mss \to \infty$, experiencing a root at some value of $\mss$.
However, there is no clear pattern regarding the positions of these roots.
From asymptotic considerations (\cf\ Section~\ref{sec:FiniteNf:SolutionStrategy}), we know that all four fixed points eventually converge to $\gGN = \gTh = 0$ as $\mss\to\infty$.
The observations here naively suggest that there may be a fusion of the \fpTh\ and \fpB\ fixed points for large flavor number and momentum transfer.
With the increase of residuals for small $\Nf$ in mind, however, this merger is likely to be an artifact indicating that the solver gets stuck at two different fixed points for small and large values of $\mss$.

Hence, while the existence of the \fpB\ fixed point is well-established for large flavor numbers, it may either vanish and wander off to infinity, or diversify, potentially developing a continuum of solutions for small $\Nf$.
Clarifying the situation will again require a finer resolution of the coupling functions' momentum dependence. 

Likewise, the spectrum at the \fpB\ fixed point (Fig.~\ref{fig:GNTh:CriticalExponents:Leading}c) shows greater uncertainties compared to the \fpTh\ (let alone the \fpGN) fixed point.
The red shading for $\Nf = 1 ... 2$ is to indicate that the large residuals and the lack of convergence in the symmetric configuration cast doubts on whether the obtained functions are actual solutions.
In any case, the fixed point has two relevant directions for large $\Nf$ as in the pointlike limit and the bosonized formulation \cite{Janssen:2012pq}.
At $\Nf = 2,3$, an additional third relevant exponent occurs in the $\msu$-favored orthogonal configuration.
This could hint at further fixed points because the corresponding eigenflows must eventually approach an \IR\ limit.
Again, though, the extra fixed point might also lie at infinity.
Nevertheless, since for these small flavor numbers the residuals of the corresponding fixed point solutions begin to increase, it remains unclear whether this extra relevant direction is physical.

\section{Conclusions}
\label{sec:Conclusions}

We have studied a class of effective theories of Dirac materials in 2+1 dimensional spacetime, featuring $\Nf$ gapless (reducible) Dirac fermion flavors on the classical level with a $\mathrm U(2\Nf)$ flavor (``chiral'') symmetry group. This class contains the (irreducible) Gross-Neveu model as well as the Thirring model classically defined in terms of local four-fermion interactions. These are widely studied similar models with partly rather different long-range properties.

We have focused on the fixed-point structure of these theories, exploring the momentum dependence of the full propagator and the interaction vertices for the first time. For this, we have used the \FRG-framework and spanned the momentum dependence in terms of Mandelstam variables. In many regimes, a reduction to the $\mss$ channel turns out to be self-consistent for quantitatively addressing the fixed-point properties. Using pseudo-spectral methods, we have constructed the momentum-dependent fixed-point solutions for the couplings at the different fixed-points of the model class. These fixed points reflect the various universality classes of this set of models which serve to define UV-complete interacting quantum field theories (asymptotically safe theories). More specifically, our results confirm the existence of the Gross-Neveu universality class for all flavors as well as the Thirring universality class for a wide range of flavors. In the large-$\Nf$ limit, we are able to construct the momentum-dependent fixed-point solutions analytically. Despite the nontrivial momentum dependence which indicates the presence of an infinite set of operators at the fixed point, both models feature only one \RG-relevant direction, making the fixed points a candidate for being associated to a 2nd-order quantum phase transition potentially visible in the long-range properties. 

In fact, for the Gross-Neveu model, the fixed point can directly be linked to a chiral phase transition to a gapped phase with an Ising-type order parameter reflecting the $\mathbbm{Z}_2$-symmetry breaking by a fermion condensation for all values of $\Nf$. The model is, however, not in the Ising universality class, as gapless fermionic modes contribute near the phase transition. Our momentum-dependent results exhibit a remarkable degree of stability of the Gross-Neveu-type interactions, lending strong support to previous quantitative \FRG{} analyses based on derivative expansions and focusing on the Gross-Neveu channel. 

The Thirring model has been suggested to be connected to a chiral phase transition of \NJL-type with symmetry breaking pattern $\mathrm U(2\Nf) \to \mathrm U(\Nf) \otimes U(\Nf)$ which, however, does not occur at large $\Nf$. Whereas our momentum-dependent results also show a large degree of stability for larger flavor number, instabilities do appear towards smaller $\Nf$. This indicates that the small $\Nf$ region requires a more careful resolution of the momentum structure of the interactions than previously anticipated. In particular, a determination of the critical flavor number $\Nfcrit$ below which a strong-coupling quantum phase transition can occur is likely to be affected by this more complex structure of the Thirring model. Our explorative studies of the momentum structure in this regime indicate that $\Nfcrit$ is smaller than previously anticipated, potentially so small that chiral symmetry does not break spontaneously in this model for any integer value of $\Nf$. While a precise estimate remains difficult, our results are compatible with recent lattice simulations that account for exact chiral symmetry providing evidence for $\Nfcrit$ being near \cite{Hands:2018vrd} or, in fact, below\cite{Lenz:2019qwu} $\Nf=1$.

In addition to the free theory, the model class also contains another interacting fixed point (fixed point \fpB) having two relevant directions for large $\Nf$. This fixed point shows the largest degree of instability with respect to a variation of momentum configurations. Hence, its properties and even its existence become questionable towards smaller $\Nf$, leaving us with the Gross-Neveu and Thirring universality classes as the relevant cases in this model set of maximal flavor symmetry.

\section*{Acknowledgements}
We would like to thank Julian Lenz, Björn Wellegehausen, and Andreas Wipf for discussions and for informing us about the results of Ref. \onlinecite{Lenz:2019qwu} prior to publication. This work was supported by the DFG-Research Training Group ``Quantum- and Gravitational Fields'' GRK 1523/2.
LD acknowledges funding by the Stiftung der Deutschen Wirtschaft.
HG acknowledges funding by the DFG under grant No. Gi328/7-1.
BK acknowledges funding by the DFG under grant No. Wi 777/11, and by the Netherlands Organisation for Scientific Research (NWO) within the Foundation for Fundamental Research on Matter (FOM) grant 13VP12.

\appendix

\section{Non-analyticity of the coupling functions}
\label{app:NonanalyticCouplingFunctions}

As claimed in the main text, the non-analyticity of the coupling functions can be traced back to the threshold kernel~\eqref{eq:ThresholdKernel}.
The decisive part is the square bracket in the first line of this equation.
Parameterizing $p^2 = \mss$ and $u = \cos\sphericalangle(q, p) = (q \cdot p) / \sqrt{q^2 p^2}$, we can expand this term around $\mss = 0$ and obtain
\begin{equation}
\label{eq:ThresholdKernel:Expansion:s0}
\begin{aligned}
	&\frac{ (a_1 + a_2) q^2 + a_1 q \sqrt{\mss}{u} - a_2 q^2 u^2 }{ q^2 \left( q^2+\mss + 2 q \sqrt{\mss} u \right)} \\
		&\;\sim \left[ a_1 + (1-u^2) a_2 \right] \frac{1}{q^2} - \left[ a_1 + 2 (1-u^2) a_2 \right] u \frac{\sqrt{\mss}}{q^3} \\
		&\;\quad - \left[ (1-2u^2) a_1 + (1 - 5u^2 + 4 u^4) a_2 \right] \frac{\mss}{q^4} \\
		&\;\quad + \left[ (3-4u^2) a_1 + (4 - 12 u^2 + 8 u^4) a_2 \right] u \frac{\mss^{3/2}}{q^5} + \mathcal{O}\!\left( \mss^2 \right) .
\end{aligned}
\end{equation}
Hence the expansion necessarily involves half-integer powers of the Mandelstam variable $\mss$.
We observe that these half-integer powers always come with odd powers of the angular cosine $u$.
Nevertheless, this does not imply that the terms vanish upon integration because the series~\eqref{eq:ThresholdKernel:Expansion:s0} is not integrable term-by-term.
Rather every additional power of $\sqrt{\mss}$ in the numerator implies a complementary power of $q$ in the denominator, so that the loop integrals of the individual terms diverge beyond linear order.

Up to linear order, on the contrary, the integrals are saved by the regulator terms in the second line of Eq.~\eqref{eq:ThresholdKernel}.
Generically, the fermionic regulator shape function goes like $r(q^2) \sim 1/q$ for $q^2 \ll 1$.
The overall contribution of the regulator terms is thus of order $q^2$ for small loop momenta.
Along with the dimensional $q^2$ term from the integration measure, this removes the divergences up to the order $\mss$ term in Eq.~\eqref{eq:ThresholdKernel:Expansion:s0}.
This also explains the maximum regularity property satisfied by the numerical finite-$\Nf$ solutions and enforced in the large-$\Nf$ limit because the term of order $\sqrt{\mss}$ then indeed vanishes due to the corresponding integrand being odd in $u$.

\bibliography{gen_bib}

\end{document}